%% file: main.tex
\documentclass[aps,prl,reprint,superscriptaddress,amsmath,amssymb,longbibliography,floatfix]{revtex4-2}
\usepackage{graphicx}
\usepackage{xcolor}
\ifdefined\pdfsuppresswarningpagegroup
\fi
\usepackage{dcolumn}
\usepackage{bm}
\usepackage{placeins}
\usepackage{amsmath,amssymb,mathtools}
\usepackage{braket}
\usepackage{physics}
\usepackage{booktabs}
\usepackage{microtype}
\usepackage{threeparttable}
\usepackage{silence}
\usepackage[colorlinks=true,
            linkcolor=blue,
            citecolor=blue,
            urlcolor=blue,
            bookmarks=true,
            bookmarksopen=true]{hyperref}
\usepackage[textsize=scriptsize]{todonotes}

\usepackage[normalem]{ulem}
\newif\ifarxiv
\arxivtrue   %
\ifarxiv
  
\else
  
\fi

\begin{document}
\hyphenation{post-se-lec-tion}

\newcommand{\papertitle}{A Forward Bit Makes Error Correction Dispensable for Antidegradable Amplitude-Damping Channels}

\title{\papertitle} 

\author{Ahmed Gamal}
\email{ametwaly@nu.edu.eg}
\affiliation{School of Information Technology and Computer Science, Nile University, Giza, Egypt}

\author{Yasutaka Wada}
\email{yasutaka.wada@mi.meijigakuin.ac.jp}
\affiliation{Department of Mathematical Informatics, Meiji Gakuin University, Kanagawa, Japan}

\author{Walid Gomaa}
\email{walid.gomaa@ejust.edu.eg}
\affiliation{Department of Computer Science and Engineering, Egypt-Japan University of Science and Technology, Alexandria, Egypt}
\affiliation{Faculty of Engineering, Alexandria University, Alexandria, Egypt}

\author{Norhan El-Sayed}
\affiliation{School of Information Technology and Computer Science, Nile University, Giza, Egypt}

\author{Keiji Kimura}
\affiliation{Department of Computer Science and Engineering, Waseda University, Tokyo, Japan}

\author{Salem F. Hegazy}
\email{sfhegazy@nu.edu.eg}
\affiliation{School of Information Technology and Computer Science, Nile University, Giza, Egypt}
\affiliation{National Institute of Laser Enhanced Sciences, Cairo University, Giza, Egypt}

\author{Kazunori Ueda}
\affiliation{Department of Computer Science and Engineering, Waseda University, Tokyo, Japan}

\author{Ahmed El-Mahdy}
\email{aelmahdy@nu.edu.eg}
\affiliation{School of Information Technology and Computer Science, Nile University, Giza, Egypt}
\affiliation{Faculty of Engineering, Alexandria University, Alexandria, Egypt}
\date{\today}

\begin{abstract}

Preserving a qubit usually relies on redundant encoding over many uses of a noisy channel. We prove that for antidegradable amplitude damping, one channel use with one forward classical bit matches or beats every forward-assisted or unassisted code with any number of uses. %
We then show that variational circuits learn to switch from error correction to classical assistance as damping grows, %
beating a dual-certified bound on all unassisted three-use codes and, relative to the best unassisted see-saw, extending storage time by up to $13\%$ at entanglement fidelity down to $0.65$. %

\end{abstract}

\maketitle

Current noisy intermediate-scale quantum (NISQ) devices suffer from limited numbers of qubits and high error rates~\cite{Preskill2018}, making it difficult to preserve quantum states for long times. Such preservation, or quantum memory, is essential for quantum repeaters~\cite{Briegel1998}, which allow long-distance quantum communication~\cite{Sangouard2011}, including quantum key distribution between distant parties~\cite{bennett2014quantum}, and distributed quantum computing~\cite{Cirac1999,Caleffi2024}. It is also needed in quantum sensing, where the sensitivity is limited by how long the sensor keeps its quantum state~\cite{degen2017quantum}, and in dynamic circuits, where qubits must wait while gates are applied depending on mid-circuit measurement outcomes~\cite{Zhou2000,Corcoles2021,Baumer2024}.

Conventional quantum error correction preserves a logical qubit by encoding it redundantly across several physical qubits, without classical communication between the encoder and the decoder~\cite{NielsenChuang2010}. In forward classical assistance~\cite{LeungMatthews2015}, a classical bit obtained from a mid-circuit measurement is transferred alongside the processed qubits. Although forward classical communication cannot increase the quantum capacity of the channel~\cite{BDSW1996}, it was shown to improve the quantum fidelity for single channel use~\cite{KorotkovKeane2010,Wang2014,Chen2017,Harraz2022}.

Because both redundant encoding and forward classical assistance can independently help preserve quantum information against decoherence, it is natural to ask how useful redundant encoding can be with forward classical assistance. %
It should be noted that there are semidefinite-programming upper bounds on the fidelity of forward-assisted and unassisted codes~\cite{LeungMatthews2015}. However, each bound is computed separately for a given number of channel uses and is not necessarily tight; thus these bounds do not show whether additional uses can improve the achievable fidelity.

For addressing this question, we consider the amplitude-damping channel with damping probability $\gamma$, since previous work~\cite{KorotkovKeane2010,Wang2014,Chen2017,Harraz2022} have shown a forward-assistance advantage by exploiting the channel bias to degrade into $\ket{0}$. Moreover, other related work has successfully provided channel specific codes~\cite{Leung1997,Fletcher2008}. It remains to show the utility of redundant coding in addition to forward assist codes.

In this Letter, we show that for an antidegradable amplitude-damping channel ($\gamma\ge1/2$), forward classical assistance is \textit{solely} able to achieve optimal fidelity with no need for redundant coding. In particular, based on the monogamy of entanglement~\cite{Kaur2019,AhmedSmith2026}, we prove an upper bound on the achievable fidelity after forward-classical-assisted channel that holds for any number of channel uses.  
This bound is also found to match the optimum fidelity of one-use channel in~\cite{Pal2014}. In the degradable region ($\gamma<1/2$), by contrast, we demonstrate that the redundant coding is able to improve the fidelity, while the forward classical assistance almost adds nothing at the weak damping region.

\begin{figure}[!tb]
\centering
\includegraphics[width=\columnwidth]{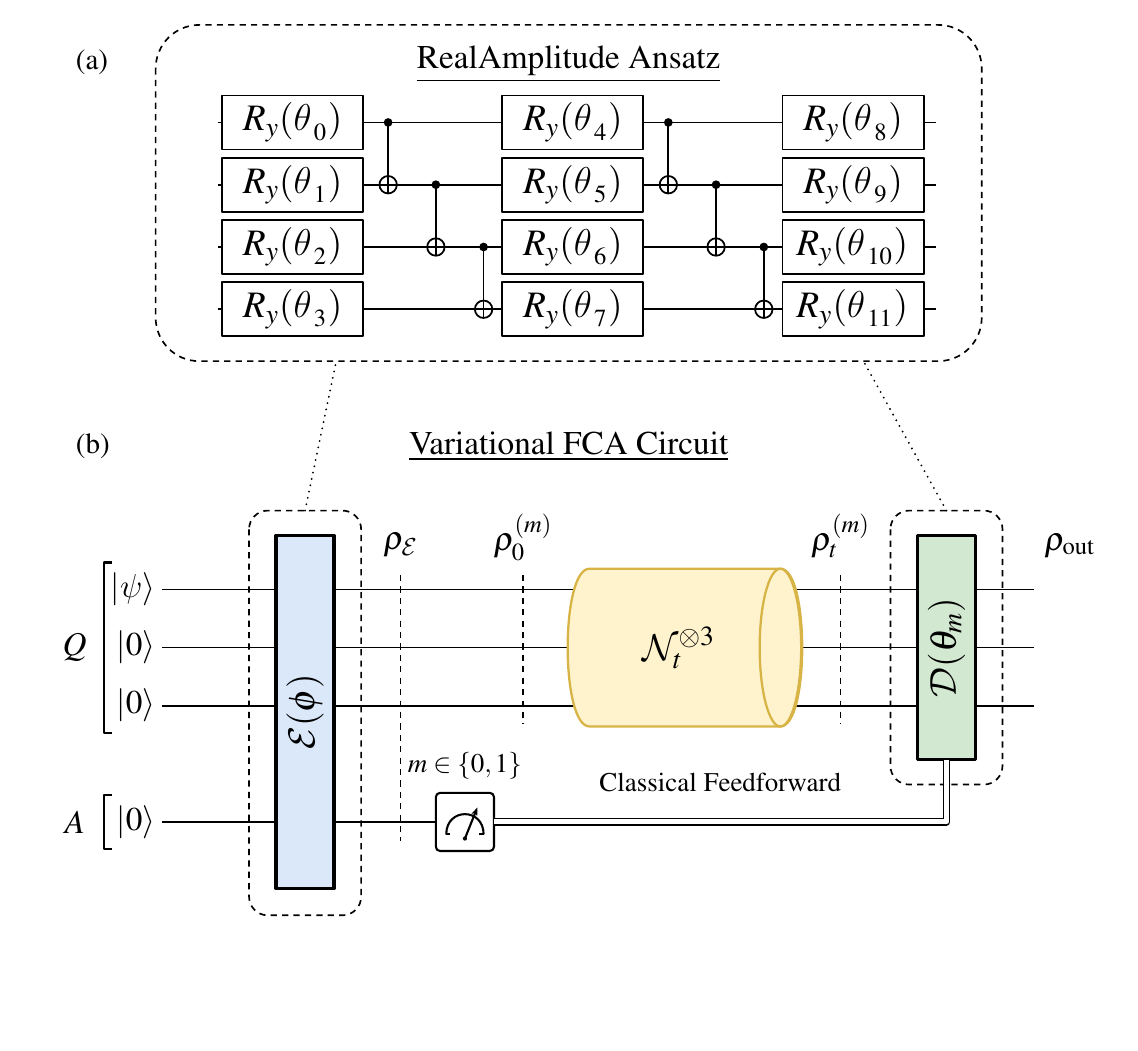}
\caption{\textbf{Forward-classically-assisted (FCA) quantum memory.}
(a) Real-amplitude ansatz used for the encoder and decoders
(b) The auxiliary qubit $A$ is measured after encoding, and only the memory qubits $Q$ are stored; the measurement outcome $m$ selects the decoder. $\rho_{\mathcal E}$, $\rho_0^{(m)}$, and $\rho_t^{(m)}$ are the states after encoding, measurement, and storage.}
\label{fig:arch}
\end{figure}

The best error-correcting scheme therefore changes with the noise level; it is redundant coding at weak damping (degradable region), and a single channel use with forward classical assistance at strong damping (antidegradable region). Having established this limit, we test its practicality by examining whether parameterized quantum circuits can learn this switch in strategy without being told. Variational circuits have previously learned quantum error-correcting codes~\cite{QVECTOR,Xu2021Compiler}, and also protocols based on local operations and classical communication (LOCC)~\cite{LOCCNet2021}, but, to our knowledge, not codes with forward classical assistance.

Figure~\ref{fig:arch} depicts a demonstration for the forward-classical-assisted approach. A variational circuit encodes a logical qubit $|\psi\rangle$ into three memory qubits, and also an auxiliary qubit that is measured with an outcome $m$ selecting one of two variational recovery circuits.

Remarkably, the trained circuit was found to reproduce the optimal transition. At weak damping, it spreads the logical qubit over all three memory qubits, whereas at strong damping it effectively uses a single memory qubit together with the classical bit and approach the one-use optimum~\cite{Pal2014}.
The output fidelity closely follows a see-saw benchmark, obtained by alternately optimizing the encoding and recovery operations~\cite{ReimpellWerner2005,KosutLidar2009}. At strong damping, it also outperforms a numerically verified upper bound on every unassisted three-use code~\cite{LeungMatthews2015}. 
This bound still lies above the fidelity that measuring and re-preparing the qubit can reach~\cite{MassarPopescu1995}, so the advantage cannot come from classical communication alone. Compared with the unassisted see-saw benchmark, the trained circuits also extend the storage time at a fixed error tolerance by up to $13\%$.

\textit{Proposed system}---
To realize forward classical assistance within a variational memory, we extend the setting of the quantum variational error corrector (QVECTOR)~\cite{QVECTOR} by an auxiliary qubit whose measurement outcome is forwarded to the decoder. Figure~\ref{fig:arch}(b) shows the resulting forward-classically-assisted (FCA) memory, which stores a logical qubit in three parallel uses of the noisy channel, one for each of three memory qubits. In particular, a four-qubit encoder $\mathcal E(\phi)$ acts on the input qubit $\rho=\ketbra{\psi}{\psi}$ together with three qubits initialized in $\ket{000}$, producing the state $\rho_{\mathcal E}=\mathcal E(\phi)(\rho\otimes\ketbra{000}{000})\mathcal E(\phi)^\dagger$. Three of these qubits form the memory register $Q$, while the fourth serves as the auxiliary qubit $A$.

The auxiliary qubit is then measured in the computational basis, giving an outcome $m\in\{0,1\}$ with probability $p(m|\rho)$. We write $\rho_0^{(m)}=\mathcal E_m(\rho)=\mathrm{Tr}_A[\ketbra{m}{m}_A\,\rho_{\mathcal E}]$ for the resulting state of $Q$ weighted by this probability, so that $\mathrm{Tr}\,\rho_0^{(m)}=p(m|\rho)$ and $Q$ is in the state $\rho_0^{(m)}/p(m|\rho)$. Only the memory qubits are stored, and each of them undergoes the amplitude-damping channel $\mathcal N_t$ for a storage time $t$, as specified in the noise model below. The register is thus left in the state $\rho_t^{(m)}=\mathcal N_t^{\otimes3}(\rho_0^{(m)})$.

The outcome $m$ is forwarded through a noiseless classical channel and selects the decoder unitary $D(\theta_m)$ acting on the three memory qubits. We keep the first qubit as the logical output and trace out the other two, which turns the decoder into the recovery map $\mathcal D_m(\sigma)=\mathrm{Tr}_{Q_2Q_3}[D(\theta_m)\,\sigma\,D(\theta_m)^\dagger]$. The output state is therefore $\rho_{\mathrm{out}}=\sum_m\mathcal D_m(\rho_t^{(m)})=\Phi_{\mathrm{eff}}^{(t)}(\rho)$, where
\begin{equation}
\Phi_{\mathrm{eff}}^{(t)}
=\sum_{m=0}^{1}
\mathcal D_m\circ\mathcal N_t^{\otimes3}\circ\mathcal E_m
\end{equation}
is the effective channel of the memory. It is worth noting that both outcomes, $m=0$ and $m=1$, contribute to $\rho_{\mathrm{out}}$ and that no run is discarded, so the memory is deterministic and requires no post-selection.

For a fair comparison, our unassisted (UA) baseline follows the register structure of QVECTOR~\cite{QVECTOR}. 
It encodes the input into three qubits, stores them in the same three parallel channel uses, and adds two fresh qubits in $\ket{00}$ before a five-qubit decoder. Both memories use the real-amplitude ansatz of Fig.~\ref{fig:arch}(a) (Supplemental Material), which has $(r+1)N_q$ $R_y$ rotations and $r(N_q-1)$ controlled-NOT (CNOT) gates for $r$ repetitions on $N_q$ qubits. Specifically, FCA uses a four-qubit encoder with $r=5$ and two three-qubit decoders with $r=7$, while UA uses a three-qubit encoder and a five-qubit decoder, both with $r=5$ (Supplemental Material). Consequently, both memories execute 48 $R_y$ rotations per run, with 29 CNOT gates for FCA and 30 for UA. Although FCA stores 72 trainable parameters, compared with 48 for UA, only the decoder selected by $m$ is executed in each run.

We quantify the performance of a memory by its fidelity averaged over Haar-random input states,
\begin{equation}
\bar{F}(t)
=\int d\psi\,
\bra{\psi}
\Phi_{\mathrm{eff}}^{(t)}
\!\left(\ketbra{\psi}{\psi}\right)
\ket{\psi}.
\label{eq:fbar}
\end{equation}
Our proof and the bounds we compare with are, however, stated in terms of the entanglement fidelity, which measures how well the memory preserves entanglement with an untouched reference qubit $R$. 
With the input and $R$ in $\ket{\Phi^+}=(\ket{00}+\ket{11})/\sqrt2$, it reads $F_e=\bra{\Phi^+}(\Phi_{\mathrm{eff}}^{(t)}\otimes\mathcal I_R)(\ketbra{\Phi^+}{\Phi^+})\ket{\Phi^+}$, and for a qubit it is related to the average fidelity by $\bar F=(2F_e+1)/3$~\cite{Horodecki1999}.

\textit{Noise model}---
To isolate the effect of the storage noise, we take encoding, decoding, measurement, and classical feedforward to be ideal, so that noise acts only during storage. In particular, each stored qubit independently undergoes the amplitude-damping channel $\mathcal N_t(\sigma)=K_0\sigma K_0^\dagger+K_1\sigma K_1^\dagger$, with Kraus operators $K_0=\ketbra{0}{0}+\sqrt{1-\gamma(t)}\,\ketbra{1}{1}$ and $K_1=\sqrt{\gamma(t)}\,\ketbra{0}{1}$~\cite{NielsenChuang2010}. Here, $\gamma(t)=1-e^{-t/T_1}$ is the probability that the excited state decays during a storage time $t$ (Supplemental Material), and we set $T_1\simeq133.1\,\mu\mathrm{s}$, the median relaxation time of the IBM processor \texttt{ibm\_fez} (Supplemental Material).

\textit{Training and evaluation}---
To obtain statistically meaningful results, we train 30 FCA circuits and 30 UA circuits from independent random initializations at each sampled storage time (listed in the Supplemental Material), using the same procedure for both. In particular, each circuit minimizes the loss $1-\bar F$ of Eq.~\eqref{eq:fbar}, estimated on 25,000 Haar-random single-qubit states (Supplemental Material). After training, we evaluate every circuit on the six Pauli eigenstates, since their equal-weight average equals the Haar average of Eq.~\eqref{eq:fbar} for any qubit channel~\cite{Bowdrey2002}. We then report the mean fidelity over the 30 circuits together with its 95\% confidence interval, of half-width $1.96\,s/\sqrt{30}$ for a sample standard deviation $s$.

\textit{Analysis}---
Our proof, given in the Supplemental Material, shows that for $\gamma\geq1/2$ no FCA or UA code, with any number of channel uses, can exceed the best one-use memory, which therefore serves as our reference. Pal \textit{et al.}~\cite{Pal2014} showed that, when one use of a qubit channel is combined with trace-preserving LOCC, the largest singlet fraction that can be shared is $\lambda_{\max}$, the largest eigenvalue of the Choi state $(\mathcal N_t\otimes\mathcal I_R)(\ketbra{\Phi^+}{\Phi^+})$ of the channel. Consequently, no one-use memory with forward classical communication exceeds $\lambda_{\max}$, and a single forward bit is already enough to reach it (Supplemental Material). Since $\bar F=(2F_e+1)/3$, the optimal average fidelity of a one-use memory is
\begin{equation}
\bar F_1=\frac{2\lambda_{\max}+1}{3}.
\label{eq:F1}
\end{equation}
For amplitude damping, $\lambda_{\max}=1-\gamma/2$, and hence $\bar F_1=1-\gamma/3$, while the condition $\gamma\geq1/2$ of the proof corresponds to $t\geq T_1\ln2\simeq92\,\mu\mathrm{s}$, where amplitude damping is antidegradable~\cite{GiovannettiFazio2005,WolfPerezGarcia2007}.

\begin{figure}[!tb]
\centering
\includegraphics[width=\columnwidth]
{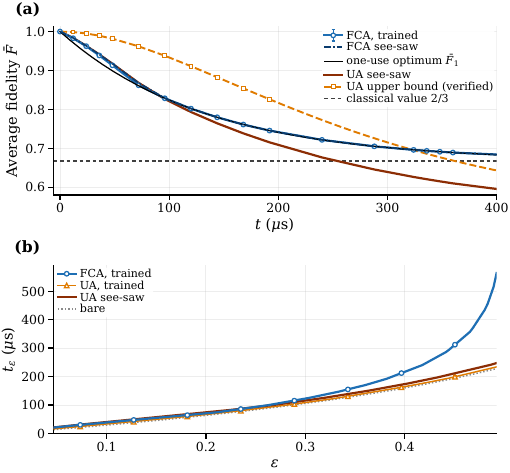}
\caption{\textbf{Trained FCA circuits follow the FCA see-saw and outperform unassisted codes.}
(a) Average fidelity versus storage time. The verified UA upper bound applies to every unassisted three-use code. The thin black line is the one-use optimum $\bar F_1$ of Eq.~\eqref{eq:F1}, and the dashed line marks the classical value $\bar F=2/3$.
(b) $\epsilon$-achievable storage time. For FCA, $t_\epsilon$ grows without bound as $\epsilon\to1/2$, because one use with the bit keeps $\bar F_1=1-\gamma/3$ above $2/3$ at every finite storage time. Crossings are interpolated linearly between sampled storage times; confidence intervals are smaller than the line widths.}
\label{fig:advantage}
\end{figure}

We next ask what this implies for unassisted codes. To bound every unassisted three-use code, we use the semidefinite-programming (SDP) relaxation of Leung and Matthews~\cite{LeungMatthews2015} that restricts codes to be non-signaling and positive-partial-transpose preserving (NS$\cap$PPTp). Each bound is certified by a dual-feasible solution, which we check numerically (Supplemental Material), and we refer to it as the verified UA upper bound. Since one forward bit alone already reaches $\bar F=2/3$ by measuring and re-preparing the input~\cite{MassarPopescu1995}, a separation is meaningful only while this bound stays above $2/3$. Computed every $6\,\mu\mathrm{s}$ between $198$ and $480\,\mu\mathrm{s}$, the bound stays above $2/3$ through $t=360\,\mu\mathrm{s}$, and, as Fig.~\ref{fig:advantage}(a) shows (thin black line and verified UA upper bound), $\bar F_1$ exceeds it from $t=324\,\mu\mathrm{s}$, by $7.8\times10^{-4}$ at $324\,\mu\mathrm{s}$ and up to $2.1\times10^{-2}$ at $360\,\mu\mathrm{s}$. In this window, one channel use together with one forward bit thus outperforms every unassisted three-use code. Moreover, with the coherence time $T_2=98.3\,\mu\mathrm{s}$ of \texttt{ibm\_fez} included, the same separation holds at $t=150$--$162\,\mu\mathrm{s}$ (Supplemental Material).

\emph{Proof sketch.}---
For rational $\gamma\in[1/2,1)$, write
$1-\gamma=k/\ell$, where $\ell\geq 2k$.
The amplitude-damping channel admits an isometric
extension to $\ell$ receivers such that any group
of $k$ receivers can reconstruct the channel
output~\cite{AhmedSmith2026}. Applying this
extension independently to all $n$ channel uses,
and providing each receiver with the classical
message, allows every group $S$ of $k$ receivers
to execute the original FCA decoder.

Let $R$ be a reference qubit maximally entangled
with the logical input. For each group $S$, define
a normalized vector $|v_S\rangle$ by projecting
the global state, $|\Psi\rangle$, onto the event that group $S$
successfully recovers a Bell pair with $R$.
Then $|\langle\Psi|v_S\rangle|^2=F_e$.
For two disjoint groups $S$ and $T$, monogamy
of entanglement implies
$|\langle v_S|v_T\rangle|\leq 1/2$.

These vectors form a $1/2$-orthonormal representation
of the Kneser graph $\mathrm{KG}_{\ell,k}$.
Applying the Kneser-graph bound of
Ref.~\cite{AhmedSmith2026} gives
\begin{equation}
F_e \leq \frac{k}{\ell}
+\frac{1}{2}\left(1-\frac{k}{\ell}\right)
=1-\frac{\gamma}{2}.
\end{equation}
Monotonicity under additional amplitude damping
extends the bound to all real $\gamma\geq1/2$.

\emph{Achievability.}---
The bound is attained using a single channel use
and one forward classical bit. Consider a two-outcome
encoding instrument with Kraus operators
\begin{equation}
M_0=\begin{pmatrix}a&0\\0&b\end{pmatrix},
\qquad
M_1=\begin{pmatrix}0&a\\b&0\end{pmatrix},
\end{equation}
where $a=1/\sqrt{2-\gamma}$ and
$b=\sqrt{1-\gamma}/\sqrt{2-\gamma}$.
The measurement outcome $m\in\{0,1\}$ is forwarded
to the decoder, which applies $U_0=I$ or $U_1=X$,
respectively, after amplitude damping.

Using the amplitude-damping Kraus operators
$K_0=\mathrm{diag}(1,\sqrt{1-\gamma})$ and
$K_1=\sqrt{\gamma}|0\rangle\langle1|$,
the entanglement fidelity is
\begin{equation}
\begin{aligned}
F_e
&=\frac14\sum_{m,j}
  |\mathrm{Tr}(U_m K_j M_m)|^2\\
&=\frac12\left(a+b\sqrt{1-\gamma}\right)^2\\
&=1-\frac{\gamma}{2}.
\end{aligned}
\end{equation}
Thus, a single channel use with one forward
classical bit achieves the universal upper bound,
so additional channel uses cannot improve the
optimal FCA fidelity for $\gamma\geq1/2$.
Full details are provided in the Supplemental
Material.

\textit{Results}---Having established these limits, we now examine whether explicit circuits can realize them, using the trained three-use FCA circuits of Fig.~\ref{fig:arch}. We compare them with the FCA see-saw, which alternately optimizes general encoding and recovery maps and places no limit on the circuit size~\cite{ReimpellWerner2005,KosutLidar2009}. Nevertheless, Fig.~\ref{fig:advantage}(a) shows that our fixed-size circuits stay within $6.1\times10^{-3}$ of the see-saw over the whole range, and within $2.3\times10^{-4}$ for $t\geq96\,\mu\mathrm{s}$, where they therefore lie close to $\bar F_1$. Moreover, the full 95\% confidence interval of trained FCA lies above the UA see-saw for $t\geq96\,\mu\mathrm{s}$, and at $t=324$, $336$, $348$, and $360\,\mu\mathrm{s}$, all 30 trained circuits exceed the verified UA upper bound; at $324\,\mu\mathrm{s}$, even the lowest trained fidelity, $0.695216$, lies above the bound of $0.695104$. At these times, their entanglement fidelity, between $0.533$ and $0.544$, stays above the value $1/2$ that no measure-and-prepare channel can exceed, and the concurrence of their Choi states~\cite{Wootters1998} remains nonzero, between $0.067$ and $0.087$. Consequently, the advantage is not obtained by a purely classical memory.

We next ask how much longer the trained circuits can preserve a logical qubit. For an error tolerance $\epsilon$, we define $t_\epsilon$ as the longest storage time with $F_e\geq1-\epsilon$, or equivalently $\bar F\geq1-2\epsilon/3$~\cite{BuscemiDatta2010}. As Fig.~\ref{fig:advantage}(b) shows, trained FCA stores the qubit longer than the UA see-saw for $\epsilon\gtrsim0.245$, with an improvement that grows from $6\%$ at $\epsilon=0.30$ to $13\%$ at $\epsilon=0.35$.

To test the role of the bit directly, we keep the FCA architecture, discard the measurement outcome, and retrain a single decoder in place of the two conditional ones (Supplemental Material). This ablation executes the same gates as FCA, but at $t=624\,\mu\mathrm{s}$ it gives $\bar F=0.5382\pm0.0006$, close to the UA see-saw value $0.5401$ and well below the $0.6697$ of trained FCA. Consequently, the strong-damping advantage of the trained circuits relies on the measurement outcome.

\begin{figure}[!tb]
\centering
\includegraphics[width=\columnwidth]
{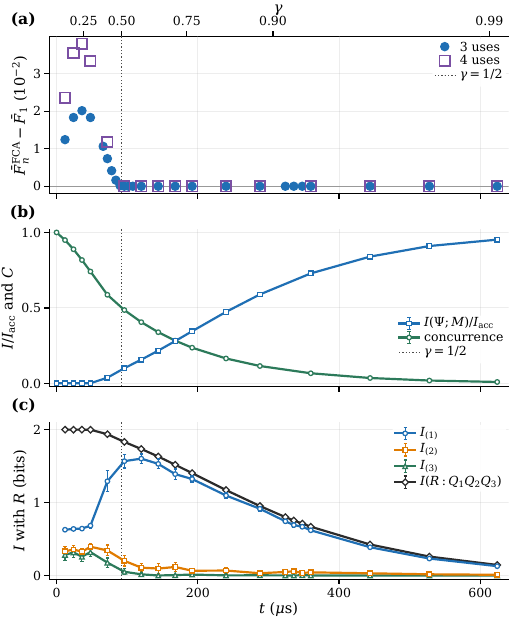}
\caption{\textbf{Extra channel uses help at weak damping; the classical bit at strong damping.}
(a) Gain of the best $n$-use see-saw over the one-use optimum $\bar F_1$. For each $n$, we plot the larger of the FCA and UA see-saws, since every UA code is also an FCA code.
(b) Mutual information in the measurement record of the trained three-use FCA circuit, normalized by $I_{\mathrm{acc}}$, and the Choi-state concurrence $C$.
(c) Quantum mutual information between a reference $R$, maximally entangled with the logical input, and the memory qubits right after encoding. $I_{(1)}\geq I_{(2)}\geq I_{(3)}$ are the single-qubit values sorted within each measurement branch and averaged with weights $p(m)$, and $I(R:Q_1Q_2Q_3)$ is the whole register. In (b) and (c), points are means over 30 trained circuits, with 95\% confidence intervals.
The dotted line marks $\gamma=1/2$ ($t\simeq92\,\mu\mathrm{s}$), beyond which amplitude damping is antidegradable and no FCA code can exceed $\bar F_1$ (Supplemental Material).}
\label{fig:mechanism}
\end{figure}

Finally, we ask whether redundant encoding still helps below $\gamma=1/2$, where our proof leaves room for it, and whether the circuits learn the predicted switch. Figure~\ref{fig:mechanism}(a) compares $\bar F_1$ with the best three- and four-use see-saws. At $t=36\,\mu\mathrm{s}$, they exceed the one-use optimum by $2.0\times10^{-2}$ and $3.8\times10^{-2}$, respectively, whereas for $t\geq96\,\mu\mathrm{s}$ they differ from it by at most $7\times10^{-7}$ and $1.5\times10^{-5}$, consistent with the proof. The gain from the bit behaves in the opposite way; relative to the UA see-saw with the same number of uses, it stays below $10^{-8}$ up to $t=72\,\mu\mathrm{s}$ and grows to about $0.13$ at $624\,\mu\mathrm{s}$ for both three and four uses (Supplemental Material). The see-saws thus show the switch from redundant encoding to the bit near $\gamma=1/2$.

The trained circuits make the same switch. To show this, we compute the mutual information $I(\Psi;M)$ between the Haar-random input $\Psi$ and the outcome $M$, estimated from 5000 inputs, and compare it with $I_{\mathrm{acc}}=1-1/(2\ln2)\simeq0.279$ bits, the largest information that any measurement can obtain about a Haar-random qubit~\cite{Barnett2010}. Figure~\ref{fig:mechanism}(b) shows that $I(\Psi;M)<2\times10^{-4}$ bits for $t\leq48\,\mu\mathrm{s}$, whereas from about $t=72\,\mu\mathrm{s}$ ($\gamma=0.42$) it grows, reaching $0.2653\pm0.0006$ bits, or $95.2\%$ of $I_{\mathrm{acc}}$, at $624\,\mu\mathrm{s}$, while the concurrence $C$ falls to $0.0090\pm0.0001$. Moreover, Fig.~\ref{fig:mechanism}(c) shows the quantum mutual information $I(R:Q_i)=S(R)+S(Q_i)-S(RQ_i)$, with $S$ the von Neumann entropy, between $R$ and each memory qubit right after encoding (Supplemental Material). Up to $t=48\,\mu\mathrm{s}$, the register holds the maximal two bits, but no single qubit holds more than about $0.67$ bits, so the logical qubit is spread as in a redundant code. From about $96\,\mu\mathrm{s}$, in contrast, one memory qubit per branch holds nearly all of it, as in the one-bit protocol of the Supplemental Material.

In summary, we have shown that a single forward classical bit makes redundant encoding unnecessary for antidegradable amplitude damping. In particular, for $\gamma\geq1/2$, one channel use together with the bit already reaches the highest fidelity of any FCA or UA code, regardless of the number of uses. Moreover, we have shown that variational circuits learn when to switch between the two strategies, relying on quantum redundancy at weak damping and on the bit at strong damping, while closely following see-saw optimization. At $t=324$--$360\,\mu\mathrm{s}$, the trained circuits outperform a verified upper bound on every unassisted three-use code while this bound is still above $2/3$, and, compared with the best unassisted code found, they extend the storage time at a fixed error tolerance by up to $13\%$. Finally, our analysis assumes ideal encoding, decoding, measurement, and classical feedforward, and including their errors and latency, as well as other noise models, is an important next step.

\textit{Acknowledgments}---The authors used Claude Fable 5.1 (Anthropic), with high reasoning effort, to assist with debugging of the code and mathematical derivations. The authors directed this use, verified all derivations and text produced with its help, and take full responsibility for all results presented in this Letter.

\FloatBarrier
\bibliography{quantum_memory_prl}
\ifarxiv
\clearpage
\onecolumngrid
\setcounter{equation}{0}
\renewcommand{\theequation}{S\arabic{equation}}
\setcounter{figure}{0}
\renewcommand{\thefigure}{S\arabic{figure}}
\setcounter{table}{0}
\renewcommand{\thetable}{S\arabic{table}}
\begin{center}
{\large Supplemental Material}\\[4pt]
\papertitle\\[6pt]
\end{center}
\vspace{0.5\baselineskip}
\input{supplement_body}
\fi
\end{document}

%% file: supplement_body.tex
We use the abbreviations of the Letter: forward-classically-assisted (FCA) and unassisted (UA) codes, local operations and classical communication (LOCC), semidefinite programming (SDP), and the controlled-NOT (CNOT) gate.

This Supplemental Material provides the technical details behind the results of the Letter. Section~S1 defines the unassisted baseline and the FCA ablation, and Sec.~S2 gives the ansatz, resource counts, and training protocol. Section~S3 describes the storage-noise model and the bare-qubit reference, and Sec.~S4 the see-saw optimization and the SDP upper bounds, including the numerical dual checks and the two-use bound. Section~S5 derives the one-use optimum and compares it with the multi-use memories, and Sec.~S6 extends the separation to amplitude and phase damping. Section~S7 defines the classical-information and entanglement diagnostics, and Secs.~S8 and S9 report trainability and output-purity diagnostics. The proof that additional channel uses cannot exceed the one-use optimum for $\gamma\geq1/2$ is given in Sec.~S5.2, and unassisted codes that approach this optimum are given in Sec.~S5.5.

\phantomsection
\section*{S1. Unassisted Baseline and FCA Ablation}
\label{app:baselines}
This section defines the unassisted (UA) baseline and the FCA ablation used in the Letter, both of which are trained with the same protocol as FCA (Sec.~S2), with 30 independent models at each storage time.
    
\subsection*{1. UA baseline}
\label{app:ua}

The UA baseline follows the register structure of the quantum variational error corrector (QVECTOR)~\cite{QVECTOR}. An encoder $V(\theta)$ acts on three qubits, which then pass through $\mathcal{N}_t^{\otimes 3}$, after which two fresh qubits in $|00\rangle$ are added and a five-qubit decoder acts on all five qubits. We keep the first qubit as the logical output and trace out the other four, with no measurement and no classical feedforward at any stage.

UA differs from the original QVECTOR circuit in two ways, both chosen to match FCA. First, the encoder and decoder use the same real-amplitude ansatz as FCA, and the decoder is an independent circuit rather than one containing the inverse encoder $V^\dagger(\theta)$. Second, as in FCA, the decoder is updated twice for each encoder update~\cite{PuenteRizzi2025}. UA therefore uses the same number of noisy channel uses, ansatz family, number of executed rotations (Sec.~S2), and training as FCA, and lacks only the measurement, the forward bit, and the second decoder.

\subsection*{2. FCA ablation}
\label{app:unconditioned}

\begin{figure}[tbp]
\centering
\includegraphics[width=0.72\linewidth]
{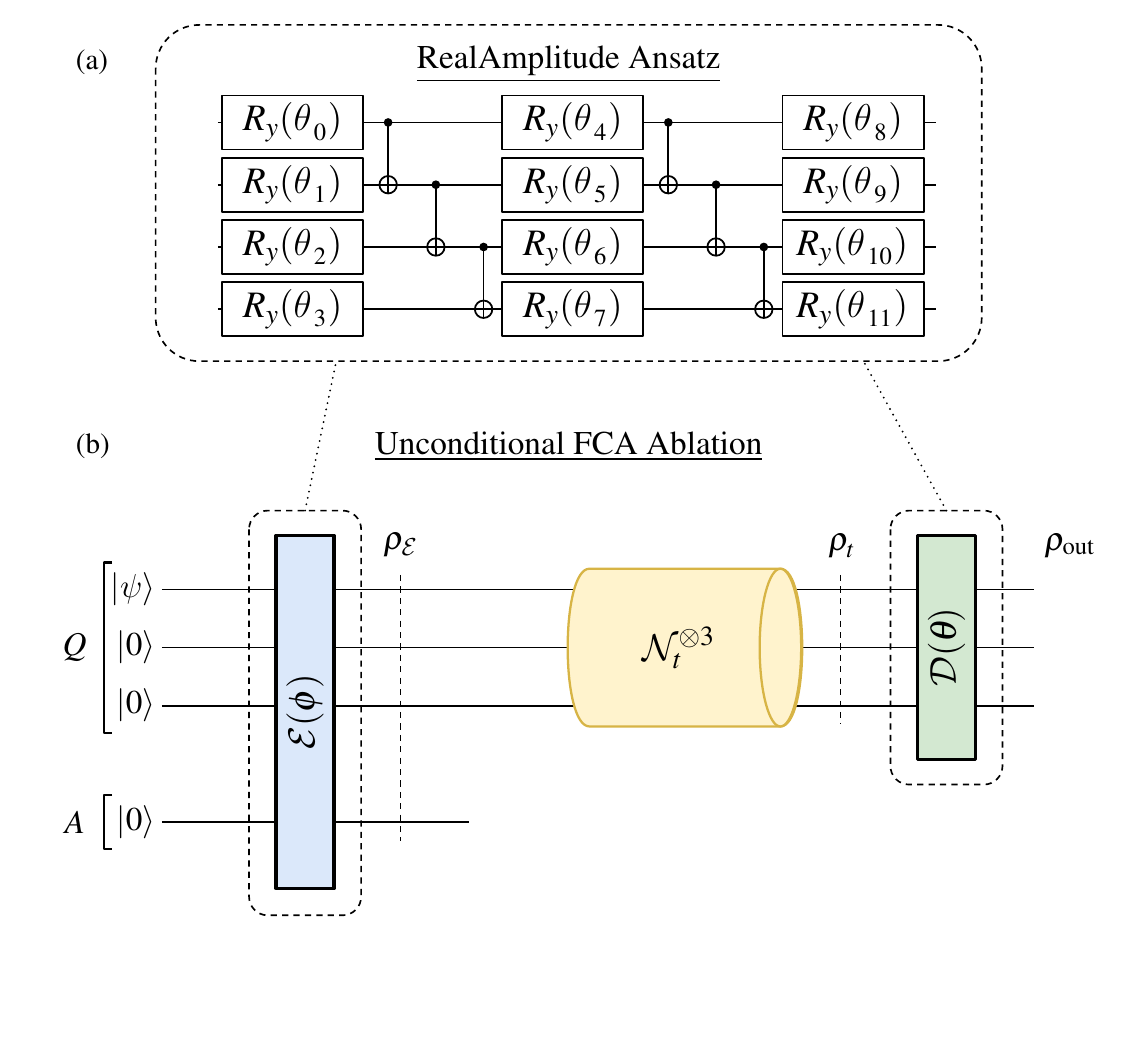}
\caption{\textbf{FCA ablation.}
(a) Real-amplitude ansatz, as in Fig.~1(a) of the Letter.
(b) The four-qubit encoder prepares the memory register $Q$ and the auxiliary qubit $A$. The auxiliary qubit is discarded, and $Q$ passes through the noisy channel. A single three-qubit decoder $\mathcal{D}(\theta)$ acts on $Q$ for every input, with no measurement record or classical feedforward. $\rho_{\mathcal E}$ and $\rho_t$ are the states after encoding and storage.}
\label{figS:unconditioned}
\end{figure}

The ablation keeps the FCA circuit but removes the classical bit. Its four-qubit encoder is the same as in FCA, but after encoding the auxiliary qubit is traced out, which is equivalent to measuring it and discarding the outcome. The three memory qubits then pass through $\mathcal{N}_t^{\otimes 3}$, and a single three-qubit decoder of the same size as one FCA decoder recovers the logical qubit (Fig.~\ref{figS:unconditioned}). The ablation therefore executes the same 48 $R_y$ rotations and 29 CNOT gates as FCA (Sec.~S2) and differs from it only in that its decoder cannot depend on the measurement outcome. We train it from scratch at each storage time, using the same protocol as FCA.

\begin{figure}[!htbp]
\centering
\includegraphics[width=0.72\linewidth]
{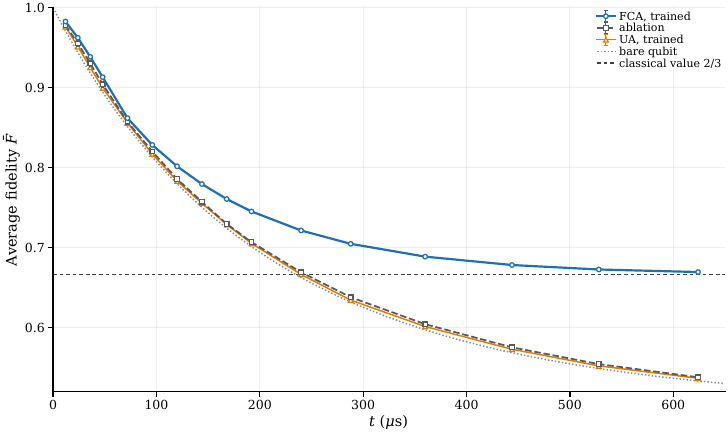}
\caption{\textbf{Effect of the forward classical bit.}
Mean Haar-averaged fidelity of trained FCA, the ablation, trained UA, and the analytic bare qubit. The dashed line marks the classical value $\bar F=2/3$. Each point is the mean over 30 independently trained models, evaluated using the six Pauli eigenstates. Error bars show 95\% confidence intervals.}
\label{figS:ablation_results}
\end{figure}

Figure~\ref{figS:ablation_results} compares the trained FCA, ablation, and UA memories. Without the measurement outcome, the code falls back to the unassisted level, since at every storage time the ablation lies within $0.006$ of trained UA and below the UA see-saw. The gap between FCA and the ablation, in contrast, is at most $0.009$ for $t\leq48\,\mu\mathrm{s}$ and grows steadily to $0.131$ at $624\,\mu\mathrm{s}$.

The small gap at weak damping does not come from the bit. For $t\leq48\,\mu\mathrm{s}$, the bit carries almost no information about the input (Sec.~S7), and trained FCA does not exceed the UA see-saw. Instead, the trained ablation and UA circuits fall short of the UA see-saw by $0.005$ to $0.018$ for $t\leq72\,\mu\mathrm{s}$, while trained FCA stays within $6.1\times10^{-3}$ of its own see-saw, so the weak-damping gap mainly reflects how well each circuit trains. From $t=96\,\mu\mathrm{s}$ on, trained FCA exceeds the UA see-saw, and the growing gap to the ablation is the advantage given by the classical bit.

\FloatBarrier

\phantomsection
\section*{S2. Ansatz and Parameter Accounting}
\label{app:ansatz_params}
FCA, UA, and the ablation use the same hardware-efficient real-amplitude ansatz family, shown in Fig.~1(a) of the Letter. Each repetition consists of an $R_y$ rotation layer followed by a linear CNOT layer, and a final $R_y$ layer follows the last repetition. Only the $R_y$ angles are trainable. An ansatz on $N_q$ qubits with $r$ repetitions therefore has $(r+1)N_q$ trainable parameters and $r(N_q-1)$ CNOT gates.

\subsection*{1. Resource counts}
Table~\ref{tabS:resources} lists the circuits used in the Letter. In FCA, the auxiliary qubit is measured after encoding, so each of the two conditional decoders acts on the three memory qubits. In UA, two fresh qubits in $\ket{00}$ are added after storage, so the decoder acts on five qubits. The ablation uses the FCA encoder and a single decoder of the same size as one FCA decoder. Only one decoder is executed in each run, so all three memories execute 48 $R_y$ rotations per run, with 29 CNOT gates for FCA and the ablation and 30 for UA. FCA stores two decoders and therefore has 72 trained parameters, compared with 48 for UA and the ablation. FCA differs from the ablation only in that it uses the measurement outcome to select which decoder is executed.

\begin{table}[!htbp]
\centering
\caption{Circuit sizes. Each memory executes its encoder and one decoder per run.}
\label{tabS:resources}
\begin{tabular}{llcccc}
\toprule
Memory & Circuit & Qubits & Repetitions & Parameters & CNOT gates\\
\midrule
FCA      & Encoder            & 4 & 5 & 24      & 15\\
         & Decoder ($\times2$) & 3 & 7 & 24 each & 14 each\\
UA       & Encoder            & 3 & 5 & 18      & 10\\
         & Decoder            & 5 & 5 & 30      & 20\\
Ablation & Encoder            & 4 & 5 & 24      & 15\\
         & Decoder            & 3 & 7 & 24      & 14\\
\bottomrule
\end{tabular}
\end{table}

\subsection*{2. Training protocol}
FCA, UA, and the ablation use the same training procedure. The training set contains 25,000 Haar-random single-qubit states, is fixed throughout each run, and is reshuffled at each epoch and processed in minibatches of 16. The training loss is therefore a sampled estimate of the Haar-averaged infidelity $1-\bar F$ of Eq.~(2) of the Letter. After training, we evaluate $\bar F$ exactly with the six Pauli eigenstates, whose equal-weight average equals the Haar average for any qubit channel~\cite{Bowdrey2002}.

We use Adam for both the encoder and the decoder parameters, with the same learning-rate schedule for both optimizers, in which the learning rate increases linearly from zero to $3\times10^{-3}$ during the first 10\% of the schedule and then follows a cosine decay to $10^{-5}$. Each optimizer evaluates the schedule with its own update counter. Because the decoder is updated twice per minibatch and the encoder only once, the decoder advances through the schedule twice as fast and reaches the $10^{-5}$ floor about halfway through training. All reported training runs use the same update rule and schedule.

For each minibatch, we perform one encoder update followed by two decoder updates. In FCA, both conditional decoders enter each decoder update, weighted by their measurement-branch probabilities, so no measurement outcomes are sampled during training.

Before training, we generate 100 candidate initializations with angles drawn uniformly from $[0,4\pi)$, evaluate each on a fixed subset of 1024 training states, and keep the one with the lowest loss. Training then runs for 15 epochs without early stopping. The FCA implementation also contains an optional measurement-probability regularizer, whose coefficient is set to zero for all results reported here.

Throughout, 95\% confidence intervals are $\pm1.96\,s/\sqrt{N_{\rm m}}$, where $s$ is the sample standard deviation over the $N_{\rm m}$ independently trained models ($N_{\rm m}=30$, or $N_{\rm m}=5$ for the four-use FCA cohort from $96\,\mu\mathrm{s}$).

\phantomsection 
\section*{S3. Storage Noise Model}
\label{app:noise_model}
We model storage as amplitude damping acting independently on each memory qubit. Only the storage is noisy. Encoding, decoding, measurement, and classical feedforward are ideal.

\subsection*{1. Amplitude damping}

For damping probability $\gamma$, the Kraus operators are~\cite{NielsenChuang2010}
\begin{equation}
K_0=
\begin{pmatrix}
1 & 0\\
0 & \sqrt{1-\gamma}
\end{pmatrix},
\qquad
K_1=
\begin{pmatrix}
0 & \sqrt{\gamma}\\
0 & 0
\end{pmatrix}.
\end{equation}
For an input state $\rho$ with elements $\rho_{ij}$, the channel gives
\begin{equation}
\mathcal N_\gamma(\rho)
=\sum_{i=0}^{1}K_i\rho K_i^\dagger
=
\begin{pmatrix}
\rho_{00}+\gamma\rho_{11} & \sqrt{1-\gamma}\,\rho_{01}\\
\sqrt{1-\gamma}\,\rho_{10} & (1-\gamma)\rho_{11}
\end{pmatrix}.
\label{eqS:ad_channel}
\end{equation}
Amplitude damping therefore lowers both the excited-state population and the coherences, and it leaves the ground state $\ket{0}$ unchanged.

\subsection*{2. Device-calibrated parameters}

We divide storage into steps of $\tau=24\,\mathrm{ns}$, the single-qubit gate duration of \texttt{ibm\_fez}, so that $L$ steps correspond to a storage time $t=L\tau$. With the median \texttt{ibm\_fez} relaxation time $T_1=133.07\,\mu\mathrm{s}$, the damping probability per step is $1-e^{-\tau/T_1}\simeq1.8034\times10^{-4}$. We round this to $\gamma_1=1.803\times10^{-4}$ in all calculations. The rounded value corresponds to $T_1\simeq133.10\,\mu\mathrm{s}$ and changes the analytic fidelities by at most $3\times10^{-5}$. Because the steps combine exactly (Sec.~S3.3), $\gamma(t)=1-e^{-t/T_1}$ does not depend on the choice of $\tau$. Training, see-saw optimization, SDP bounds, and analytic references all use this same channel.

\subsection*{3. Accumulation over storage steps}

After $L$ steps, the excited-state population is multiplied by $(1-\gamma_1)^L$. The $L$ steps therefore combine exactly into a single amplitude-damping channel, Eq.~\eqref{eqS:ad_channel}, with
\begin{equation}
\gamma(t)=1-(1-\gamma_1)^{t/\tau}.
\end{equation}
This is the channel $\mathcal N_t\equiv\mathcal N_{\gamma(t)}$ of the Letter, and the three memory qubits undergo $\mathcal N_t^{\otimes3}$. We therefore evaluate each storage time with this one channel, and no sampling over noise realizations is needed.

\subsection*{4. Storage times}

We train each architecture separately at each storage time, so every circuit is optimized for the noise strength at which it is evaluated. The main grid is $t=0$, $12$, $24$, $36$, $48$, $72$, $96$, $120$, $144$, $168$, $192$, $240$, $288$, $360$, $444$, $528$, and $624\,\mu\mathrm{s}$. UA is trained on this grid, and the ablation on the same grid without $t=0$. We also train FCA at $t=324$, $336$, and $348\,\mu\mathrm{s}$ to resolve the window $t=324$--$360\,\mu\mathrm{s}$, in which the verified UA upper bound lies below the one-use optimum $\bar F_1$ but still above $2/3$. The see-saw fidelities and the verified UA upper bound are computed at every FCA training time with $t>0$. At $t=0$ the channel is the identity, and every fidelity equals 1. The bound is also computed every $6\,\mu\mathrm{s}$ between $198$ and $480\,\mu\mathrm{s}$ (Sec.~S4).

\subsection*{5. Bare-qubit reference}
We also compare with a single unencoded qubit stored in the same channel. The channel maps a Bloch vector $\mathbf r$ to $\big(\sqrt{1-\gamma}\,r_x,\ \sqrt{1-\gamma}\,r_y,\ (1-\gamma)r_z+\gamma\big)$, and averaging the pure-state fidelity $(1+\mathbf r\cdot\mathbf r')/2$ over Haar-random inputs gives
\begin{equation}
\bar F_{\mathrm{bare}}(t)=\frac12+\frac{\sqrt{1-\gamma(t)}}{3}+\frac{1-\gamma(t)}{6}.
\label{eqS:bare}
\end{equation}
Since this expression is analytic, the bare-qubit curve has no sampling uncertainty.

\phantomsection
\section*{S4. See-Saw Optimization and SDP Bounds}
\label{app:sdp_bounds}
The trained circuits are restricted to the ansatz of Sec.~S2. To separate their performance from the limits of each coding class, we use two further calculations. The see-saw optimization gives an achievable fidelity using general encoder and decoder maps, while SDP relaxations give upper bounds.

\subsection*{1. See-saw optimization}
\label{app:seesaw}

The encoder and decoder cannot be optimized jointly as a convex problem. If one is fixed, the fidelity is linear in the Choi matrix of the other. We therefore alternate encoder and decoder updates~\cite{ReimpellWerner2005,KosutLidar2009}. Each subproblem is an SDP, but instead of calling a solver we update the Choi matrix with a fixed-point iteration, with at most 20,000 iterations per inner update and a stopping tolerance of $10^{-15}$.

For FCA, the encoder is a general completely positive and trace-preserving (CPTP) map from the logical qubit to four qubits. The fourth qubit is then measured in the computational basis, and the outcome selects one of two general CPTP decoders acting on the three stored qubits. Because the encoder is general, fixing the measurement basis loses no generality. In the implementation, each decoder also receives the measured qubit reset to $\ket{0}$. Since this input is fixed, the decoders are still general CPTP maps on the three stored qubits. Both outcomes contribute to the average fidelity.

For UA, the encoder is a general CPTP map from the logical qubit to three qubits, which pass through $\mathcal N_t^{\otimes3}$, followed by a general CPTP decoder. There is no measurement or classical feedforward. The four-use FCA and UA see-saws and the one-use see-saw (Sec.~S5) are defined in the same way.

Each sweep returns a valid protocol, so every reported see-saw fidelity is achievable and is a lower bound on the optimal fidelity of its class. The optimization does not establish a global optimum.

All restarts begin from random maps with fixed random-number seeds. We use 100 restarts at each storage time for the three-use FCA and UA see-saws, 64 for the one-use see-saw, and 50 for the additional three-use FCA and UA runs every $6\,\mu\mathrm{s}$ between $66$ and $108\,\mu\mathrm{s}$. The four-use FCA and UA see-saws continue the best of 96 restarts. A run stops when one full sweep improves the fidelity by less than $10^{-12}$, or when it reaches a sweep limit. This limit is 200 sweeps for three-use FCA and UA, raised to 1000 at $12$, $48$, $72$, and $192\,\mu\mathrm{s}$ for FCA and at $t\leq144\,\mu\mathrm{s}$ for UA. It is 1000 for the four-use and the additional three-use runs, and 60 for the one-use see-saw. We report the largest fidelity over the restarts.  

In a few cases the best run reached the sweep limit. This happened at 1 of 19 storage times for three-use FCA and for UA (both at $12\,\mu\mathrm{s}$), at 2 of 16 for four-use FCA ($12$ and $72\,\mu\mathrm{s}$), at 1 of 16 for four-use UA ($12\,\mu\mathrm{s}$), at 6 of 19 for the one-use see-saw (all with $t\leq96\,\mu\mathrm{s}$), and at 3 and 1 of the 6 additional FCA and UA times ($78$, $84$, $90\,\mu\mathrm{s}$ and $102\,\mu\mathrm{s}$). These values still come from valid protocols, but they may lie slightly below the converged see-saw value. For the one-use see-saw, where the exact value is known, the shortfall is at most $1.15\times10^{-5}$ (Sec.~S5). None of the certified separations in the Letter depends on see-saw convergence, because each one compares a trained circuit or the one-use optimum $\bar F_1$ with the verified UA upper bound.

\subsection*{2. SDP upper bounds}
\label{app:ppt_bounds}

To upper-bound the optimal fidelity, we use the semidefinite coding relaxations of Leung and Matthews~\cite{LeungMatthews2015}. These relaxations replace the physical encoder--decoder protocol by a larger class of operations that can be optimized as an SDP. We use their relaxations for codes that are non-signaling and positive-partial-transpose-preserving (NS$\cap$PPTp) and for PPT-preserving codes (PPTp).

Let $A'$ and $B$ denote the three-qubit input and output spaces of the storage channel, so that $d_{A'}=d_B=8$. We use the unnormalized Choi operator
\begin{equation}
J_t=\left(\operatorname{id}_{A'}\otimes\mathcal N_t^{\otimes3}\right)\left(\ketbra{\Omega_8}{\Omega_8}\right),
\qquad
\ket{\Omega_8}=\sum_{j=0}^{7}\ket{j}_{A'}\ket{j}_B,
\label{eqS:channel_choi}
\end{equation}
so that $\Tr J_t=8$ and $\Tr_B J_t=I_{A'}$. This differs from the normalized Choi state used for the one-use optimum in Sec.~S5. For one logical qubit, the logical dimension is $K=2$, and the SDP maximizes the entanglement fidelity $F_e=\Tr[J_t^{T}\Lambda]$, where the transpose is the full matrix transpose in the $A'_1A'_2A'_3B_1B_2B_3$ basis. The optimization variables are $\rho$ on $A'$, the average channel input state, and $\Lambda$ on $A'\otimes B$, which represents the code in the relaxation. They satisfy
\begin{equation}
\Lambda\succeq0,\qquad \Lambda\preceq\rho\otimes I_B,\qquad \rho\succeq0,\qquad \Tr\rho=1.
\label{eqS:sdp_base}
\end{equation}
The non-signaling condition is $\Tr_{A'}\Lambda=I_B/K^2$, and the PPT-preserving condition, with the partial transpose $T_B$ on the output $B$, is
\begin{equation}
-\frac{\rho\otimes I_B}{K}\leq T_B(\Lambda)\leq\frac{\rho\otimes I_B}{K}.
\label{eqS:ppt_constraint}
\end{equation}
Every unassisted code lies in the NS$\cap$PPTp class, so $F_e^{\mathrm{UA}}\leq F_e^{\mathrm{NS}\cap\mathrm{PPTp}}$. FCA allows forward classical communication, so we do not impose the non-signaling constraint. Every FCA code is PPT-preserving, so $F_e^{\mathrm{FCA}}\leq F_e^{\mathrm{PPTp}}$. The PPTp relaxation is larger than the one-bit FCA class, so its value is an outer bound rather than the optimum of that class, and in this work we use it only for the one-use check in Sec.~S5. We convert all SDP values to average fidelity with $\bar F=(2F_e+1)/3$.
\subsection*{3. Numerical evaluation}
\label{app:sdp_numerics}
We first check that the see-saw maps define physical channels. The fixed-point update uses an inverse square root whose eigenvalues are regularized at $10^{-14}$. In addition, the FCA decoders never receive the reset qubit in $\ket{1}$, so their action on that input is not fixed by the optimization. Both effects can leave decoders that are not exactly trace preserving. We therefore complete the missing trace with a fixed output state, remove negative Choi eigenvalues, and renormalize to restore trace preservation. We then check complete positivity and trace preservation again.

These checks cover all 101 see-saw protocols, namely three-use FCA and UA at 19 storage times, four-use FCA and UA at 16, one-use at 19, and the additional three-use FCA and UA runs at 6. The smallest repaired Choi eigenvalue is $-1.60\times10^{-15}$, within round-off, and the largest trace-preservation residual is $1.33\times10^{-14}$ in Frobenius norm. The repair changes the average fidelity by at most $9.8\times10^{-8}$. We recompute each fidelity by propagating the six Pauli eigenstates through the full repaired protocol. These values agree with the see-saw objective to within $6.7\times10^{-16}$. All reported see-saw fidelities use these checked values.

The upper-bound SDPs are implemented in CVXPY and solved using \textsc{Clarabel}. We compute both the NS$\cap$PPTp and PPTp bounds at 60 storage times. These are the 19 sampled times with $t>0$ together with every $6\,\mu\mathrm{s}$ between $198$ and $480\,\mu\mathrm{s}$, where 7 times belong to both sets.

A numerical solver can return small violations of the dual constraints. We therefore repair and check the dual solution at every storage time. We first symmetrize the dual matrices and remove negative eigenvalues from those required to be positive semidefinite. We then add small identity shifts to the appropriate dual variable and increase the scalar dual variable as needed, targeting a $10^{-9}$ margin in the remaining matrix inequalities. We recompute all constraint residuals after the repair. A feasible dual solution gives an upper bound by weak duality~\cite{LeungMatthews2015}. The feasibility checks use double-precision arithmetic, not interval arithmetic.

This procedure is applied to all 60 calculations of each bound. In every case, the repaired dual value lies above the numerical primal value, and all dual constraints hold with a margin of at least $10^{-9}$. For NS$\cap$PPTp, the largest change in the average-fidelity bound is $3.4\times10^{-8}$ and the largest identity shift is $4.4\times10^{-9}$. For PPTp, the corresponding values are $5.8\times10^{-8}$ and $6.1\times10^{-9}$. The three-use PPTp bound lies well above the three-use FCA see-saw, so it cannot decide whether extra channel uses help. We therefore do not use it in the Letter. We refer to the dual-verified NS$\cap$PPTp bound as the verified UA upper bound.

On the dense grid, the verified UA upper bound first falls below the one-use optimum $\bar F_1=1-\gamma/3$ (Sec.~S5) at $t=324\,\mu\mathrm{s}$, where it is $0.695104$ compared with $0.695887$. At the previous grid point, $318\,\mu\mathrm{s}$, the bound is still higher by $2.7\times10^{-3}$. The bound stays above $2/3$ up to $360\,\mu\mathrm{s}$, where it is $0.668435$, and falls below $2/3$ at $366\,\mu\mathrm{s}$. Thus, between $324$ and $360\,\mu\mathrm{s}$, one channel use with one forward classical bit (Sec.~S5) outperforms every unassisted three-use code, even though the bound is still above the classical value $2/3$.

\subsection*{4. Two-use UA upper bound}
\label{app:twouse}
We also compute the NS$\cap$PPTp bound for unassisted codes with two channel uses, every $6\,\mu\mathrm{s}$ between $198$ and $480\,\mu\mathrm{s}$, with the same dual repair and checks as in Sec.~S4.3. This shows how the separation changes when the unassisted code has one channel use fewer. At all 48 storage times, the repaired dual value lies above the primal value, and all dual constraints hold with a margin of at least $10^{-9}$. The largest change in the bound from the repair is $2.1\times10^{-8}$, and the largest identity shift is $5.1\times10^{-9}$.

The one-use optimum $\bar F_1$ (Sec.~S5) exceeds this bound from $t=234\,\mu\mathrm{s}$ ($\gamma=0.828$), by $1.1\times10^{-3}$. The bound stays above $2/3$ up to $306\,\mu\mathrm{s}$ ($\gamma=0.900$), where it is $0.667555$ and the margin has grown to $3.3\times10^{-2}$. Thus, between $234$ and $306\,\mu\mathrm{s}$, one channel use with one forward classical bit outperforms every unassisted two-use code. With three unassisted uses, the window is $324$--$360\,\mu\mathrm{s}$. The more channel uses the unassisted code has, the stronger the damping must be for classical communication to win. Section~S5.5 shows that unassisted codes approach $\bar F_1$ as the number of uses grows, with a gap of at most $\gamma^{n-1}/3$.

\phantomsection
\section*{S5. One-Use Optimum}
\label{app:oneuse}

To separate the gain from the classical bit from the gain from quantum redundancy, we compare the three- and four-use memories with the best memory that uses the channel only once. This section derives the fidelity of that memory and compares it with the multi-use memories.

\subsection*{1. Optimum for one channel use}
\label{app:oneuse_exact}

The normalized Choi state of one use of the storage channel is
\begin{equation}
\rho_{\mathcal N_t}=(\mathrm{id}\otimes\mathcal N_t)\left(\ketbra{\Phi^+}{\Phi^+}\right),
\qquad
\ket{\Phi^+}=\frac{\ket{00}+\ket{11}}{\sqrt2}.
\end{equation}
For amplitude damping, with $\eta=1-\gamma(t)$,
\begin{equation}
\rho_{\mathcal N_t}=\frac12\Big[\ketbra{00}{00}+\sqrt{\eta}\left(\ketbra{00}{11}+\ketbra{11}{00}\right)+\eta\ketbra{11}{11}+\gamma\ketbra{10}{10}\Big],
\end{equation}
whose largest eigenvalue is $\lambda_{\max}=(1+\eta)/2=1-\gamma/2$.

Pal \textit{et al.} showed that $\lambda_{\max}$ is the largest singlet fraction that can be obtained by sending one half of a two-qubit state through one channel use and then applying trace-preserving LOCC~\cite{Pal2014}. A one-use memory with forward classical communication is a protocol of this kind. To see this, we apply the memory to one half of $\ket{\Phi^+}$. For each outcome $m$, the sender passes one half of a two-qubit state through $\mathcal N_t$ and sends $m$ forward, and the receiver applies the decoder for $m$. Each branch is therefore a protocol of the type considered by Pal \textit{et al.}, and so is their average. The overlap of the resulting state with $\ket{\Phi^+}$ is the entanglement fidelity $F_e$ of the memory, so $F_e\leq\lambda_{\max}$. For a qubit channel, $\bar F=(2F_e+1)/3$~\cite{Horodecki1999}, and every one-use memory with forward classical communication satisfies
\begin{equation}
\bar F\leq\bar F_1=\frac{2\lambda_{\max}+1}{3}=1-\frac{\gamma}{3}.
\label{eqS:pal_oneuse}
\end{equation}

Pal \textit{et al.}~\cite{Pal2014} showed that $\lambda_{\max}$ is reached in entanglement sharing, where the sender prepares a two-qubit state and sends one half of it through the channel. For amplitude damping, the optimal state to send is $\ket{\chi}=a\ket{00}+b\ket{11}$, with $a=1/\sqrt{1+\eta}$ and $b=\sqrt{\eta}/\sqrt{1+\eta}$, where $\eta=1-\gamma$ is the probability that $\ket{1}$ does not decay~\cite{BandyopadhyayGhosh2012,Pal2014}. For $\gamma>0$, this state is not maximally entangled, and no processing after the channel is needed. A memory, however, cannot choose the state it stores, since its logical qubit is maximally entangled with the reference in $\ket{\Phi^+}$. Moreover, for nonunital channels such as amplitude damping, sending one half of a maximally entangled state and then applying any trace-preserving LOCC gives a singlet fraction strictly below $\lambda_{\max}$ (Theorem~3 of Ref.~\cite{Pal2014}). A memory that reaches $\lambda_{\max}$ must therefore act on the logical qubit before storage. We now show that one forward classical bit is enough for this.

Before storage, the sender applies the two-outcome instrument
\begin{equation}
M_0=
\begin{pmatrix}
a&0\\
0&b
\end{pmatrix},
\qquad
M_1=
\begin{pmatrix}
0&a\\
b&0
\end{pmatrix},
\label{eqS:onebit_instrument}
\end{equation}
which is valid, since $M_0^\dagger M_0+M_1^\dagger M_1=I$. With the logical qubit in $\ket{\Phi^+}$ with the reference, each outcome occurs with probability $\Tr(M_m^\dagger M_m)/2=1/2$. For $m=0$, the instrument turns $\ket{\Phi^+}$ into $\ket{\chi}$, the optimal state of Pal \textit{et al.}, while for $m=1$ it gives $(X\otimes I)\ket{\chi}$, the same state with the reference flipped, where $X$ is the Pauli bit-flip operator. The sender sends $m$ as one classical bit, and after storage the receiver applies $U_0=I$ for $m=0$ and $U_1=X$ for $m=1$. Since $(X\otimes X)\ket{\Phi^+}=\ket{\Phi^+}$, this correction compensates the flip of the reference, so that both outcomes give the fidelity of $\ket{\chi}$. The bit therefore allows the memory to store the optimal state of Pal \textit{et al.} for every outcome. This instrument has the same form as the weak measurement with pre- and post-flips of Chen \textit{et al.}~\cite{Chen2017}, who used it to protect known GHZ states. As a direct check, with the Kraus operators $K_j$ of Sec.~S3.1, the entanglement fidelity is~\cite{NielsenChuang2010}
\begin{equation}
F_e=\frac14\sum_{m=0}^{1}\sum_{j=0}^{1}\left|\Tr(U_m K_j M_m)\right|^2
=\frac12(a+\sqrt{\eta}\,b)^2=1-\frac{\gamma}{2}=\lambda_{\max},
\label{eqS:onebit_fidelity}
\end{equation}
where only the two terms without decay contribute. The corresponding average fidelity is $\bar F=(2F_e+1)/3=1-\gamma/3=\bar F_1$.

Thus, one noisy channel use and one forward classical
bit attain the one-use LOCC optimum for every
$0\leq\gamma\leq1$. The protocol is explicit and does not rely on numerical optimization.

\begin{figure}[!htbp]
\centering
\includegraphics[width=0.75\linewidth]
{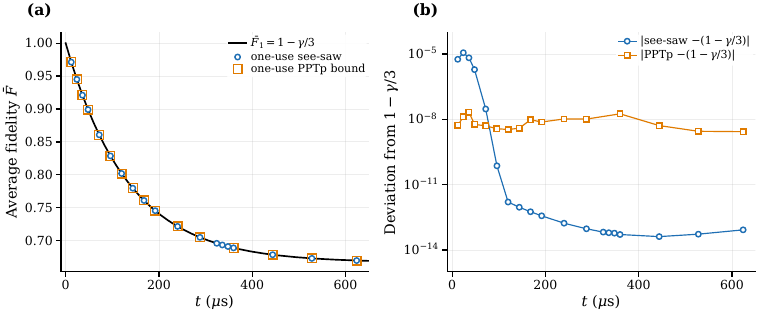}
\caption{\textbf{One-use optimum.}
(a) The one-use optimum $1-\gamma/3$ (line), checked one-use see-saw fidelities, and verified one-use PPTp upper bounds.
(b) Absolute differences from $1-\gamma/3$ on a logarithmic scale. The see-saw lies within $1.15\times10^{-5}$ of $1-\gamma/3$ at every storage time, and the PPTp bound within $2.2\times10^{-8}$.}
\label{figS:oneuse_validation}
\end{figure}

As a numerical check, the verified one-use PPTp
bound (Sec.~S4) lies above $\bar F_1$ by at most
$2.2\times10^{-8}$ at the tested storage times,
while the one-use see-saw falls short of $\bar F_1$
by at most $1.15\times10^{-5}$
(Fig.~\ref{figS:oneuse_validation}).

Without classical communication, storing the qubit directly is already the best one-use memory. The verified one-use NS$\cap$PPTp bound exceeds $\bar F_{\mathrm{bare}}$ of Eq.~\eqref{eqS:bare} by at most $8.3\times10^{-9}$ at all 19 storage times, so no unassisted one-use encoder and decoder improves on the bare qubit. The one-use optimum therefore exceeds every unassisted one-use memory by
\begin{equation}
\bar F_1-\bar F_{\mathrm{bare}}=\frac{(1-\sqrt{\eta})^2}{6}>0
\qquad\text{for every }\gamma>0.
\end{equation}
With one channel use, forward classical assistance therefore always helps. Section~S5.2 shows that, for $\gamma\geq1/2$, redundant encoding over more channel uses cannot improve on this one-use value.

\subsection*{2. Bound for any number of channel uses}
\label{app:anyuse}
\input{proof}

\subsection*{3. Gain from additional channel uses}
\label{app:oneuse_redundancy}
For $n$ channel uses, we define
\begin{equation}
\Delta F_n(t)=\bar F_{\mathrm{ss}}^{(n)}(t)-\bar F_1(t),
\label{eqS:redundancy_gain}
\end{equation}
where $\bar F_{\mathrm{ss}}^{(n)}$ is the larger of the $n$-use FCA and UA see-saws, since every UA code is also an FCA code. Since $\bar F_1$ bounds every one-use memory with classical communication, a positive $\Delta F_n$ is a gain that no one-use memory can reach. For three uses, $\Delta F_3$ is largest at $t=36\,\mu\mathrm{s}$, where it is $2.0\times10^{-2}$. From $t=96\,\mu\mathrm{s}$, $|\Delta F_3|\leq7\times10^{-7}$, and between $324$ and $360\,\mu\mathrm{s}$, $|\Delta F_3|\leq1.1\times10^{-9}$. For four uses, $\Delta F_4$ is also largest at $36\,\mu\mathrm{s}$, where it is $3.8\times10^{-2}$, and from $96\,\mu\mathrm{s}$, $|\Delta F_4|\leq1.5\times10^{-5}$. No value from $96\,\mu\mathrm{s}$ is positive, and the small negative values come from incomplete see-saw convergence. Extra channel uses therefore help at weak damping, while from about $96\,\mu\mathrm{s}$ the see-saws find no gain over one use. The transition lies between $72\,\mu\mathrm{s}$ ($\gamma=0.42$) and $96\,\mu\mathrm{s}$ ($\gamma=0.51$), around $\gamma=1/2$ ($t=T_1\ln2\simeq92.3\,\mu\mathrm{s}$), where amplitude damping becomes antidegradable and its quantum capacity vanishes~\cite{GiovannettiFazio2005,WolfPerezGarcia2007}. For $\gamma\geq1/2$, Sec.~S5.2 proves that no FCA code, with any number of channel uses, exceeds $\bar F_1$, so from $96\,\mu\mathrm{s}$ the see-saws reach the optimum to within the deviations above. For $\gamma<1/2$, the see-saw gives the gain found, not the largest possible gain.

The Letter also compares each FCA see-saw with the UA see-saw with the same number of uses. For three uses, we add see-saw runs every $6\,\mu\mathrm{s}$ between $66$ and $108\,\mu\mathrm{s}$ (Sec.~S4). Up to $72\,\mu\mathrm{s}$, the best three-use FCA code found is an unassisted code, and the four-use FCA and UA see-saws agree to within $10^{-12}$. From $96\,\mu\mathrm{s}$, the FCA see-saws agree with $\bar F_1$ and lie above the UA see-saws, by $0.13$ at $624\,\mu\mathrm{s}$. By linear interpolation, the three-use UA see-saw crosses $\bar F_1$ at $89\,\mu\mathrm{s}$ ($\gamma=0.49$). Only near this crossing does the three-use FCA see-saw exceed both references, by $1.5\times10^{-6}$, $2.0\times10^{-5}$, and $1.5\times10^{-4}$ at $78$, $84$, and $90\,\mu\mathrm{s}$. The largest difference probably lies near the crossing, between grid points. At these three times the FCA runs reached the sweep limit, so longer runs could only increase these values.

\subsection*{4. Four-use FCA}
\label{app:fouruse}

The four-use FCA memory uses five qubits. The encoder acts on all five qubits with 5 repetitions, giving 30 parameters and 20 CNOT gates. One qubit is measured, and each of the two decoders acts on the four stored qubits with 7 repetitions, giving 32 parameters and 21 CNOT gates. Each run uses the encoder and one decoder, that is, 62 parameters and 41 CNOT gates, and 94 parameters are stored in total.

The four-use circuit is trained with the protocol of Sec.~S2 at 16 storage times, from $12$ to $624\,\mu\mathrm{s}$. For $t\leq72\,\mu\mathrm{s}$ we train 30 models for 15 epochs, at $96\,\mu\mathrm{s}$ 5 models for 15 epochs, and for $t\geq120\,\mu\mathrm{s}$ 5 models for 8 epochs. The trained circuit falls furthest below its see-saw at $36\,\mu\mathrm{s}$, by $3.5\times10^{-2}$. At this time 30 models were trained for the full 15 epochs, so the shortfall is not caused by the shorter training used at stronger damping. It may instead come from the restricted $R_y$ ansatz or from the initialization, and a more expressive ansatz, for example with general single-qubit rotations, may be needed. At $120\,\mu\mathrm{s}$, four of the five models lie within $6\times10^{-5}$ of $1-\gamma/3$, but one is stuck at $0.780$ and lowers the mean. Because only 5 models are used for $t\geq96\,\mu\mathrm{s}$, the trained four-use results there serve only as a qualitative check.

\subsection*{5. Unassisted codes that approach the one-use value}
\label{app:ua_approach}
Unassisted codes can approach the one-use value $F_e=1-\gamma/2$ by carrying the bit through the channel. The encoder applies the one-bit instrument of Sec.~S5.1 and keeps the logical qubit in one channel use, as before. Instead of sending $m$ as a classical bit, it prepares $n-1$ further qubits in $\ket{m}^{\otimes(n-1)}$, and all $n$ qubits pass through $\mathcal N_t$. The decoder measures these $n-1$ qubits, sets $\hat m=1$ if any of them reads 1 and $\hat m=0$ otherwise, and applies $U_{\hat m}$. Since $\mathcal N_t$ leaves $\ket{0}$ unchanged, $\hat m\neq m$ only if $m=1$ and all $n-1$ copies decay, which happens with probability $\gamma^{n-1}$. The outcome $m=1$ occurs with probability $1/2$ (Sec.~S5.1), so its contribution to $F_e$ is at most $1/2$, and misreading it lowers $F_e$ by at most $\gamma^{n-1}/2$. This unassisted code therefore has $F_e\geq1-\gamma/2-\gamma^{n-1}/2$. Together with the bound of Sec.~S5.2, the best unassisted $n$-use code satisfies, for $\gamma\geq\tfrac12$,
\begin{equation}
1-\frac{\gamma}{2}-\frac{\gamma^{n-1}}{2}\;\le\;F_e^{\mathrm{UA}}(n)\;\le\;1-\frac{\gamma}{2}.
\label{eq:ua_sandwich}
\end{equation}
For $\gamma<1$, as $n\to\infty$, the lower bound approaches $1-\gamma/2$, so unassisted codes approach the one-use value from below but never exceed it. The deficit $\gamma^{n-1}/2$ shrinks geometrically, so this code comes within $\delta$ of the one-use value, for $0<\delta<\tfrac12$ and $\gamma<1$, once
\begin{equation}
n\;\geq\;1+\frac{\ln(2\delta)}{\ln\gamma}.
\label{eq:n_sufficient}
\end{equation}
With $\gamma=1-e^{-t/T_1}$ and $\ln(1-x)\approx-x$ for small $x$, the number of uses that suffices for this code grows as $\ln[1/(2\delta)]\,e^{t/T_1}$, exponentially in the storage time. This is an upper bound on the number of uses needed; whether other unassisted codes approach the one-use value with fewer uses is left open.

\FloatBarrier
\phantomsection
\section*{S6. Amplitude and Phase Damping}
\label{app:apd}

\begin{figure}[!htbp]
\centering
\includegraphics[width=0.75\linewidth]
{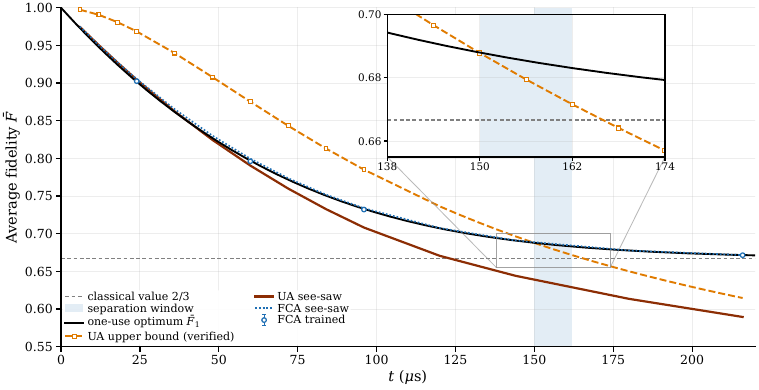}
\caption{\textbf{Separation with amplitude and phase damping.}
Average fidelity versus storage time. The solid line is the one-use optimum $\bar F_1$. Also shown are the verified UA upper bound on every unassisted three-use code (computed every $6\,\mu\mathrm{s}$ from $120\,\mu\mathrm{s}$), the UA and FCA see-saws, trained FCA at four storage times, and the classical value $\bar F=2/3$. The shaded window, $150$--$162\,\mu\mathrm{s}$, marks where $\bar F_1$ exceeds the bound while the bound is above $2/3$. Inset: zoom on the window, with the bound shown at its computed grid points.}
\label{figS:apd}
\end{figure}

We now test whether the separation survives dephasing. The test uses only quantities that do not depend on the variational circuit, namely the verified UA upper bound, the see-saws, and the one-use optimum, so it needs no new training. Trained FCA at four storage times is shown in Fig.~\ref{figS:apd} only for comparison.

\subsection*{1. Channel}

At each storage step, amplitude damping is followed by phase damping with Kraus operators $P_0=\mathrm{diag}(1,\sqrt{1-\kappa_1})$ and $P_1=\mathrm{diag}(0,\sqrt{\kappa_1})$. We set $\sqrt{1-\kappa_1}=e^{-\tau/T_\phi}$, where the dephasing time $T_\phi$ follows from $1/T_2=1/(2T_1)+1/T_\phi$. The median \texttt{ibm\_fez} value $T_2=98.32\,\mu\mathrm{s}$ gives $T_\phi\simeq155.9\,\mu\mathrm{s}$ and $\kappa_1=3.078\times10^{-4}$. At storage time $t$, $\kappa(t)=1-(1-\kappa_1)^{t/\tau}$, and the combined channel is
\begin{equation}
\mathcal N_t(\rho)=
\begin{pmatrix}
\rho_{00}+\gamma\rho_{11} & c\,\rho_{01}\\
c\,\rho_{10} & \eta\,\rho_{11}
\end{pmatrix},
\qquad
\eta=1-\gamma,
\qquad
c=\sqrt{\eta(1-\kappa)} .
\end{equation}
It has Kraus operators $K_0=\mathrm{diag}(1,c)$, $K_1=\sqrt{\eta-c^2}\,\ketbra{1}{1}$, and $K_2=\sqrt{\gamma}\,\ketbra{0}{1}$. Pure amplitude damping is the case $\kappa=0$, $c=\sqrt{\eta}$.

\subsection*{2. One-use optimum}

The normalized Choi state is
\begin{equation}
\rho_{\mathcal N_t}=\tfrac12\Big[\ketbra{00}{00}+c\left(\ketbra{00}{11}+\ketbra{11}{00}\right)+\eta\ketbra{11}{11}+\gamma\ketbra{10}{10}\Big].
\end{equation}
Its largest eigenvalue comes from the block $C=\tfrac12\begin{pmatrix}1&c\\c&\eta\end{pmatrix}$,
\begin{equation}
\lambda_{\max}=\frac{1+\eta+\sqrt{(1-\eta)^2+4c^2}}{4}.
\end{equation}
The argument of Sec.~S5.1 then gives $\bar F\leq\bar F_1=(2\lambda_{\max}+1)/3$ for every one-use memory with forward classical communication. One forward bit is again enough to reach this value. We use the instrument of Eq.~\eqref{eqS:onebit_instrument}, now with general real weights $a\geq b\geq0$ satisfying $a^2+b^2=1$, and the same corrections $U_0=I$ and $U_1=X$. With the Kraus operators above, its entanglement fidelity is
\begin{equation}
F_e=\tfrac12\left(a^2+2abc+\eta b^2\right)
=\tfrac12\begin{pmatrix}a&b\end{pmatrix}\begin{pmatrix}1&c\\c&\eta\end{pmatrix}\begin{pmatrix}a\\b\end{pmatrix},
\end{equation}
which equals $\lambda_{\max}$ when $(a,b)$ is the principal eigenvector of this matrix. For $\kappa=0$, this is the instrument of Sec.~S5.1. As a check, the one-use PPTp bound agrees with $\bar F_1$ to within $1.9\times10^{-8}$, and the one-use see-saw agrees to within $1.1\times10^{-5}$.

\subsection*{3. Separation}
We compute the verified UA upper bound for three channel uses (Sec.~S4) at 27 storage times. These are 14 times between $6$ and $216\,\mu\mathrm{s}$ together with every $6\,\mu\mathrm{s}$ between $120$ and $216\,\mu\mathrm{s}$, where 4 times belong to both sets. All duals pass the checks of Sec.~S4.3. The largest change from the dual repair is $5.7\times10^{-8}$, and the largest identity shift is $8.1\times10^{-9}$.

The one-use optimum $\bar F_1$ first exceeds the bound at $t=150\,\mu\mathrm{s}$ ($\gamma=0.676$), by $1.5\times10^{-4}$. At $144\,\mu\mathrm{s}$, the bound is still higher by $5.7\times10^{-3}$. The bound stays above $2/3$ up to $162\,\mu\mathrm{s}$ ($\gamma=0.704$) and falls below it at $168\,\mu\mathrm{s}$. Thus, between $150$ and $162\,\mu\mathrm{s}$, $\bar F_1$ exceeds the verified UA upper bound by $1.5\times10^{-4}$, $5.9\times10^{-3}$, and $1.15\times10^{-2}$, while the bound is still above $2/3$. The smallest margin is about $2.6\times10^{3}$ times the largest dual-repair correction.

In this window, the one-use optimum corresponds to an entanglement fidelity $F_e=\lambda_{\max}$ between $0.532$ and $0.525$, above the value $1/2$ that no measure-and-prepare memory can exceed, so the memory is still quantum. Compared with pure amplitude damping, the window is earlier and narrower, and the three-use see-saw gain over the one-use optimum fades more slowly, from $3.4\times10^{-4}$ at $120\,\mu\mathrm{s}$ to $2\times10^{-5}$ at $180\,\mu\mathrm{s}$. We do not report storage-time gains for this channel, because we did not train unassisted circuits for it. The $\epsilon$-achievable storage times in the Letter refer to pure amplitude damping only.

\FloatBarrier
\phantomsection
\section*{S7. Classical Side Information and Choi Entanglement}
\label{app:mechanism}

Figures~3(b) and 3(c) of the Letter use three diagnostics to study how the learned FCA memory changes with storage time. The first measures the state-dependent information carried by the measurement record, the second the entanglement retained by the learned effective quantum channel, and the third where the encoder places the logical qubit.

At each sampled storage time, we compute both diagnostics separately for each of the 30 independently trained FCA models and report the cohort mean. Error bars in Figs.~3(b) and 3(c) show 95\% confidence intervals over the 30 models (Sec.~S2.2).

\subsection*{1. Information carried by the measurement outcome}

For an input state $\rho_\psi=\ketbra{\psi}{\psi}$, the outcome $m$ occurs with probability $p(m|\psi)=\Tr[\mathcal E_m(\rho_\psi)]$, where $\mathcal E_m$ is the encoder branch for outcome $m$ (Letter, Proposed system). With the Haar average $p(m)=\int d\psi\,p(m|\psi)$, the mutual information between the input-state label $\Psi$ and the outcome $M$ is
\begin{equation}
I(\Psi;M)=\int d\psi\sum_{m=0}^{1}p(m|\psi)\log_2\frac{p(m|\psi)}{p(m)}.
\label{eqS:mutual_information}
\end{equation}
Mutual information is nonlinear in the branch probabilities, so the six-state identity used for the average fidelity does not apply. We therefore estimate the Haar integrals with $N=5000$ Haar-random pure qubit states $\psi_j$, drawn independently of the training set, using $\widehat p(m)=N^{-1}\sum_{j}p(m|\psi_j)$ and
\begin{equation}
\widehat I(\Psi;M) = \frac{1}{N} \sum_{j=1}^{N} \sum_{m=0}^{1} p(m|\psi_j) \log_2 \frac{p(m|\psi_j)}{\widehat p(m)}.
\label{eqS:mutual_information_empirical}
\end{equation}
The same sample is used for every model and storage time. We compute the mutual information separately for each encoder and then average over the models. The error bars therefore describe the variation across the 30 trained models for this fixed input sample. They do not include the Monte Carlo uncertainty of the Haar integration.

Although the measurement is performed before the storage channel, $I(\Psi;M)$ depends on the storage time $t$ because a separate encoder is trained for each sampled storage time. The learned measurement therefore changes with the storage noise that the encoder was trained for.

For a single pure qubit drawn from the Haar ensemble, the maximum mutual information obtainable by a measurement is $I_{\mathrm{acc}}=1-1/(2\ln2)\simeq0.27865$ bits~\cite{Barnett2010}. At $t=624\,\mu\mathrm{s}$,

\subsection*{2. Concurrence of the learned logical channel}

We prepare a maximally entangled pair and apply the complete learned memory channel at storage time $t$ to one qubit, while the other qubit serves as an untouched reference. The resulting normalized Choi state is
\begin{equation}
\rho_\Phi = \left( \mathrm{id}\otimes\Phi_{\mathrm{eff}}^{(t)} \right) \left( |\Phi^+\rangle\langle\Phi^+| \right).
\end{equation}

We quantify the entanglement of $\rho_\Phi$ by its concurrence $C$~\cite{Wootters1998}, which is zero for separable states and one for maximally entangled states.

We compute the concurrence separately for the complete learned effective channel of each of the 30 FCA models, including both measurement-conditioned branches. The Choi state is obtained exactly by propagating basis inputs through the circuit, with no sampling. Figure~3(b) of the Letter reports the mean concurrence over the 30-model cohort together with its $95\%$ confidence interval. The cohort mean falls from $C=0.587$ at $72\,\mu\mathrm{s}$ to $0.067$ at $360\,\mu\mathrm{s}$ and $0.0090$ at $624\,\mu\mathrm{s}$.

These diagnostics do not decompose the fidelity into classical and quantum contributions. $I(\Psi;M)$ measures the information carried by the measurement record about the input state. The concurrence measures entanglement between the reference and the recovered logical qubit. As storage noise increases, the record becomes more informative and the Choi concurrence decreases. Whether keeping the record helps the recovery is tested separately by the ablation in Sec.~S1.

\subsection*{3. Location of the logical information}
To see where the encoder places the logical qubit, we compute the quantum mutual information between the reference $R$ and the memory qubits right after encoding and measurement, before storage. We prepare $R$ and the input in $\ket{\Phi^+}$, apply the trained encoder, and project the auxiliary qubit onto the outcome $m$. This gives the normalized branch state $\rho^{(m)}$ of $RQ_1Q_2Q_3$, which occurs with probability $p(m)$. In each branch, we compute
\begin{equation}
I^{(m)}(R:Q_i)=S\big(\rho^{(m)}_R\big)+S\big(\rho^{(m)}_{Q_i}\big)-S\big(\rho^{(m)}_{RQ_i}\big),
\label{eqS:qmi}
\end{equation}
where $S$ is the von Neumann entropy in bits, and we sort the three values, $I^{(m)}_{(1)}\geq I^{(m)}_{(2)}\geq I^{(m)}_{(3)}$. Figure~3(c) of the Letter shows the weighted averages $I_{(k)}=\sum_m p(m)\,I^{(m)}_{(k)}$. Since the values are sorted within each branch, $I_{(1)}$ can belong to a different physical qubit in the two branches. For the whole register, $I(R:Q_1Q_2Q_3)=\sum_m p(m)\,I^{(m)}(R:Q_1Q_2Q_3)$. Each branch state is pure, so $I^{(m)}(R:Q_1Q_2Q_3)=2S\big(\rho^{(m)}_R\big)\leq2$ bits. This value falls below 2 bits when the outcome carries information about the input, since that information then sits in the classical record rather than in the register. All quantities are computed exactly from the trained circuits, separately for each of the 30 models, and Fig.~3(c) shows the mean with 95\% confidence intervals (Sec.~S2.2).

\phantomsection
\section*{S8. Random-Landscape Trainability Diagnostics}
\label{app:trainability}

If FCA had a much easier random landscape than UA, its advantage could reflect trainability rather than the classical bit. We test this by asking whether the three register structures, which share the same ansatz family, differ in loss variance or gradient size at random parameters. Small gradients and concentrated losses are commonly used as indicators of hard-to-train variational landscapes~\cite{McClean2018,Cerezo2021,Deshpande2025}.

We sample parameters from $U(0,4\pi)$ and evaluate the untrained loss $\mathcal{L}(\theta)=1-\bar F(\theta)$ on the same fixed subset of 1024 training states for every architecture and depth. We use $t=24\,\mu\mathrm{s}$ with the amplitude-damping channel of Sec.~S3. For FCA, both measurement outcomes are kept, as in Eq.~(1) of the Letter, so no branch is postselected.

For FCA and UA, we vary the encoder and decoder repetitions over $1\leq r_{\mathrm{enc}},r_{\mathrm{dec}}\leq10$. This gives 100 depth pairs for each, with executed parameter counts of 14--77 for FCA and 16--88 for UA. Each depth pair is evaluated with 200 independent random parameter vectors, without the best-of-100 initialization used in training. The ablation is evaluated at its Letter configuration only and is not shown in the figure. The FCA gradient includes the encoder and both conditional decoders, so at the Letter configurations it has 72 components, compared with 48 for UA and the ablation, although each run of all three circuits uses 48 parameters. Because gradient norms grow with the number of components, we also report the mean absolute gradient component.

\begin{figure}[tbp]
\centering
\includegraphics[width=0.85\linewidth]
{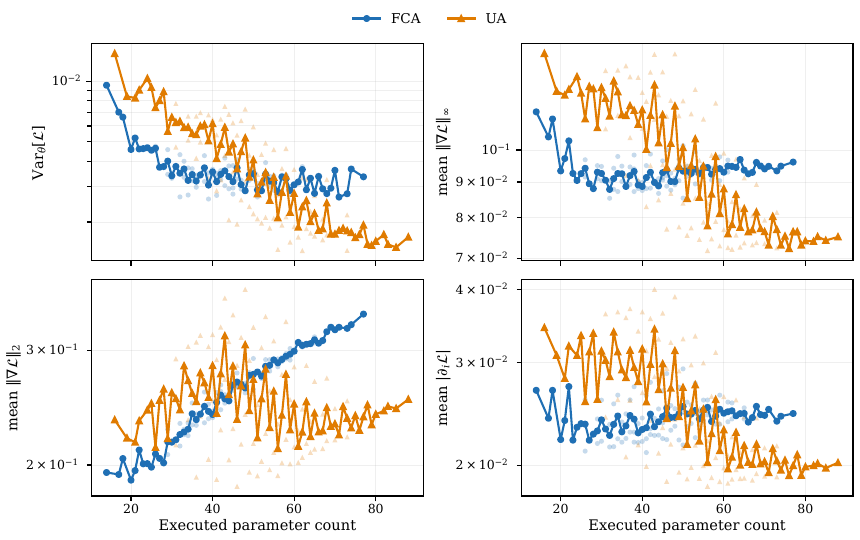}
\caption{\textbf{Random-landscape trainability diagnostics.}
FCA and UA at $t=24\,\mu\mathrm{s}$, with random parameters from $U(0,4\pi)$ and untrained loss $\mathcal{L}=1-\bar F$. The panels show $\mathrm{Var}_{\theta}[\mathcal{L}]$, the mean gradient sup norm, the mean Euclidean gradient norm, and the mean absolute gradient component versus the executed parameter count. Faint markers show individual encoder--decoder depth pairs, and solid lines average depth pairs with the same count. Values at the Letter configurations, and for the ablation, are given in the text.}
\label{figS:trainability}
\end{figure}

Figure~\ref{figS:trainability} shows the loss variance, the mean gradient sup norm, the mean Euclidean gradient norm, and the mean absolute gradient component versus the executed parameter count. Of the 53 parameter counts shared by FCA and UA, UA has the larger loss variance at 31, the larger sup norm at 33, and the larger mean absolute gradient component at 31. Averaged over the full sweep, the loss variance is $3.5\times10^{-3}$ for FCA and $4.0\times10^{-3}$ for UA. The Euclidean norm is usually larger for FCA, because its gradient has more components. At the Letter configurations, the loss variance is $2.89\times10^{-3}$ for FCA, $3.18\times10^{-3}$ for UA, and $2.90\times10^{-3}$ for the ablation, and the mean absolute gradient component is $2.33\times10^{-2}$, $2.40\times10^{-2}$, and $2.97\times10^{-2}$, respectively.

As the circuits grow, the loss variance and gradients of UA decrease, while those of FCA stay roughly constant. This is consistent with reports that mid-circuit measurements can reduce the concentration of variational landscapes~\cite{Deshpande2025}. Our test, however, keeps the register size fixed and does not establish the absence of barren plateaus.

Overall, at the Letter configurations the three random landscapes have similar loss variances and gradient sizes, and UA is not harder to train by these measures. These statistics therefore do not explain why trained FCA lies closer to its see-saw at weak damping than the trained unassisted circuits (Sec.~S1). They describe random circuits before training, so they cannot tell us which architecture is easier to optimize.

\FloatBarrier
\phantomsection
\section*{S9. Logical-Output Purity at Random Parameters}
\label{app:haar_expressibility}

A more expressive circuit could also explain the FCA advantage. We therefore ask how often random circuit parameters produce a logical output with purity above a chosen threshold, and how varied these outputs are. After decoding, we trace out the other qubits and compute the purity of the remaining logical qubit. This test measures output mixedness for a fixed input. It does not measure recovery fidelity or the expressibility of the full logical channel.

For each random parameter vector $\theta_i\sim U(0,4\pi)$, we fix the input to $\ket{0}$, run the memory circuit without storage noise, and retain the final logical qubit $\rho(\theta_i)$. For FCA, both measurement outcomes are kept, as in Eq.~(1) of the Letter, so no branch is postselected. We use the circuit configurations of the Letter. FCA and the ablation use $(r_{\mathrm{enc}},r_{\mathrm{dec}})=(5,7)$ and UA uses $(5,5)$, so each run of all three circuits uses 48 parameterized rotations.

We record the fraction of sampled outputs with purity $\Tr[\rho(\theta_i)^2]\geq0.7$. This threshold is used only for this diagnostic. It still admits mixed states and does not imply successful recovery. No samples are discarded in the memory-fidelity evaluations.

For each architecture, sampling continues until 1000 outputs satisfy the purity threshold. From the accepted outputs, we sample 50,000 pairs with replacement, excluding self-pairs. For each pair, we compute the Uhlmann--Jozsa fidelity $F(\rho_i,\rho_j)=\big[\Tr\sqrt{\sqrt{\rho_i}\,\rho_j\sqrt{\rho_i}}\big]^2$. For independent pure qubit states drawn from the Haar ensemble, the pairwise-fidelity distribution is uniform on $[0,1]$. We use this distribution as a reference for the accepted outputs. These outputs can be mixed, so the comparison does not isolate their spread over Bloch-vector directions. Because the input and all gates are real, the output density matrices stay real. Their Bloch vectors therefore lie in the $x$--$z$ plane, and no output ensemble samples the full Bloch sphere.

\begin{figure}[tbp]
\centering
\textbf{(a)}\\[-1pt]
\includegraphics[width=0.86\linewidth]
{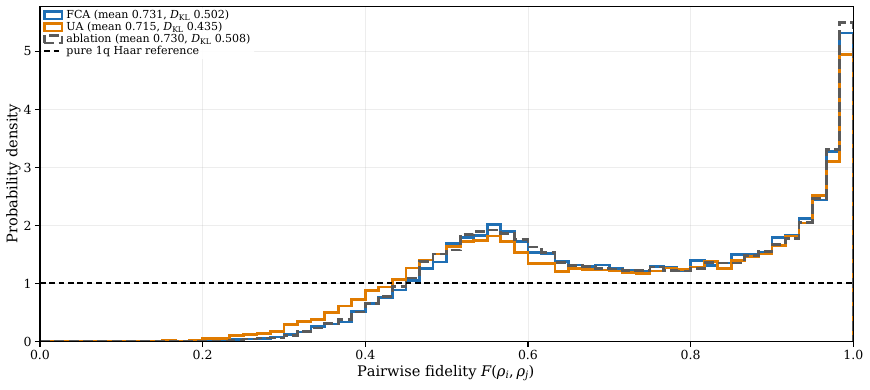}\\[4pt]
\textbf{(b)}\\[-1pt]
\includegraphics[width=0.52\linewidth]
{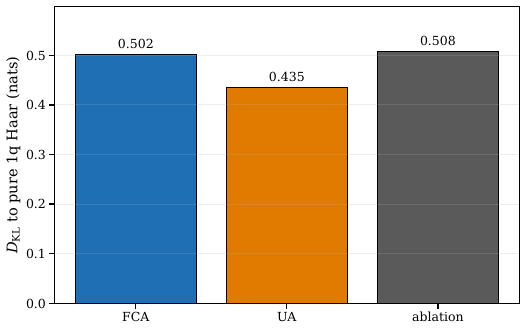}
\caption{\textbf{Pairwise similarities of purity-filtered logical outputs.}
The input is $\ket{0}$ and storage noise is absent. (a) Pairwise-fidelity distributions for outputs satisfying $\Tr(\rho^2)\geq0.7$, for FCA, UA, and the ablation. The horizontal line gives the distribution for independent pure Haar-random qubits. (b) KL divergence from this reference, in nats.}
\label{figS:purity_filtered_expressibility}
\end{figure}

We compute the Kullback--Leibler (KL) divergence $D_{\mathrm{KL}}(p_{\mathrm{emp}}\|p_{\mathrm{Haar}})$ using 60 equal-width bins on $[0,1]$. A pseudocount of $0.5$ is added to each empirical bin before normalization. We use natural logarithms, so the divergence is given in nats. Figure~\ref{figS:purity_filtered_expressibility} shows the pairwise-fidelity distributions and their KL divergences.

\begin{table}[tbp]
\centering
\caption{Purity-filtered random-output statistics. The input is $\ket{0}$ and storage noise is absent. The rate is the fraction of sampled outputs with $\Tr(\rho^2)\geq0.7$. The mean pairwise fidelity and KL divergence refer to the 1000 accepted outputs. KL divergence is reported in nats.}
\label{tabS:purity_filtered_expressibility}
\begingroup
\footnotesize
\setlength{\tabcolsep}{4pt}
\begin{tabular}{lccc}
\toprule
Architecture & Rate & $\overline{F}_{\mathrm{pair}}$ & $D_{\mathrm{KL}}$ \\
\midrule
FCA      & $18.0\%$ & $0.731$ & $0.502$ \\
UA       & $23.2\%$ & $0.715$ & $0.435$ \\
Ablation & $17.7\%$ & $0.730$ & $0.508$ \\
\bottomrule
\end{tabular}
\endgroup
\end{table}

UA produces outputs above the purity threshold more often than FCA and the ablation (Table~\ref{tabS:purity_filtered_expressibility}), and its accepted outputs have a lower mean pairwise fidelity and a smaller KL divergence from the reference. FCA and the ablation, which share the same encoder, give nearly identical statistics, so the forward bit does not change these random-parameter statistics. Because the three ensembles have different acceptance rates, these comparisons do not establish which architecture has greater logical-channel expressibility. They do not favor FCA, however, so the advantage reported in the Letter is not explained by a more expressive random circuit.

These results concern random parameters, the input $\ket{0}$, and zero storage noise. A pure output also need not equal the input, so recovery performance is assessed separately through the input--output fidelity reported in the Letter.

%% file: proof.tex
\providecommand{\ket}[1]{\lvert #1\rangle}
\providecommand{\bra}[1]{\langle #1\rvert}
\providecommand{\braket}[2]{\langle #1\vert #2\rangle}
\providecommand{\Tr}{\operatorname{Tr}}
\providecommand{\AD}{\mathcal{N}_{\gamma}}
\providecommand{\refq}{R}          %
\providecommand{\ksub}{\subseteq_{k}}
\providecommand{\KG}{\mathrm{KG}_{\ell,k}}
\providecommand{\Phip}{\Phi^{+}}
\providecommand{\Fe}{F_{e}}
\providecommand{\texorpdfstring}[2]{#1}   %

\emph{Theorem.}  For $\gamma\ge\tfrac12$, every FCA code, with any number $n$ of uses of $\AD$ and any classical message,
has entanglement fidelity $\Fe\le 1-\gamma/2$. Here, $1-\gamma/2$ is the largest eigenvalue $\lambda_{\max}$ of the Choi state of $\AD$, which Pal \emph{et al.}~\cite{Pal2014} showed to be the largest singlet fraction that one channel use with trace-preserving LOCC can give; hence, it also bounds the entanglement fidelity of any one-use FCA code. We derive this value for amplitude damping in Sec.~S5.1.

\emph{Conventions.} $\refq$ is the reference qubit that is maximally
entangled with the logical input and is left untouched, so that $\Fe$ is the
probability that the decoded qubit passes the Bell test for
$\ket{\Phip}=(\ket{00}+\ket{11})/\sqrt2$ with $\refq$.

Decoders for each group of $k$ qubits of receivers $A_i$, $0<i\le\ell$, where $i$, $k$, and $\ell$ are integers defined below, may be taken to be unitaries (append ancilla qubits), and the classical message $m$
may be kept as a quantum register in the basis $\{\ket{m}\}$, since the
decoders act on it only by operations controlled on that basis; the global
state after encoding and transmission is therefore pure.  We assume $\Fe>0$
and, until the paragraph on irrational $\gamma$, that $\gamma$ is a rational number.

\emph{Extension of the channel.}  Ahmed and Smith~\cite{AhmedSmith2026}
note that the amplitude damping channel with $\gamma=\frac{p}{p+q}$, where
$p\ge q$ are integers, is $(p,q)$-extendible: its output can be distributed
among $p+q$ receivers ($A_i$ for $0< i\le p+q$) such that any $q$ of them can recover the output of
the channel.  Put $k=q$ and $\ell=p+q$. This gives

\begin{equation}
  1-\gamma=\frac{k}{\ell},\qquad \ell\ge 2k.
  \label{eq:kn}
\end{equation}
Explicitly, the extension is the isometry
\begin{equation}
  V\ket{0}=\ket{0}^{\otimes \ell},\qquad
  V\ket{1}=\frac{1}{\sqrt{\ell}}\sum_{j=1}^{\ell}\ket{1_j},
  \label{eq:V}
\end{equation}
where $\ket{1_j}$ is the one-hot state with the $1$ in position $j$:
tracing out any $\ell-k$ of the output qubits leaves the remaining $k$
holding the output of $\AD$ (as $\ket{0}^{\otimes k}$ and
$k^{-1/2}\sum_{j\in S}\ket{1_j}$).

We apply $V$ to each of the $n$ channel uses and give receiver $j$ the
$j$-th qubit of every use together with a copy of the classical message.
  Any group $S\ksub[\ell]$ of $k$ receivers
holds the output of every channel use and the message, so it can run the
code's decoder.  Let $U_S$ be this unitary, acting only on the systems of
group $S$, let $L_S$ be its output qubit, and let $\ket{\Psi}$ be the resulting pure state of $\refq$, the $\ell$ receivers, and the encoder's environment, as well as per-group decoder ancillas $E_S$ (initialised in $\ket{0}$). Also, let

\begin{equation}
  \Pi_S=U_S^\dagger\big(\Phip_{\refq L_S}\otimes I\big)U_S
\end{equation}
be the projector ``group $S$ decodes, then the Bell test $\Phip_{\refq L_S}$
is applied'', so that the fidelity of group $S$ is
\begin{equation}
  F_S=\bra{\Psi}\Pi_S\ket{\Psi}=\Fe\qquad\text{for every }S\ksub[\ell].
  \label{eq:FS}
\end{equation}

\emph{Vectors.}  Put $\ket{\psi}=\ket{\Psi}$ and
\begin{equation}
  \ket{v_S}=\frac{\Pi_S\ket{\Psi}}{\sqrt{F_S}} .
\end{equation}
Then $\ket{v_S}$ is a unit vector with
\begin{equation}
  |\braket{\psi}{v_S}|^2=F_S .
  \label{eq:handle}
\end{equation}

\emph{Overlap of disjoint groups.}  Let $S\cap T=\emptyset$.  Then $U_S$
and $U_T$ act on different systems, so $U_TU_S$ is a unitary, and
applying it does not change inner products. Since $\ket{v_S}$ has been
projected onto $\ket{\Phip}_{\refq L_S}$ after group $S$ decoded, and $U_T$
does not touch $\refq L_S$,
\begin{align}
  U_TU_S\ket{v_S}&=\ket{\Phip}_{\refq L_S}\otimes\ket{\alpha},\nonumber\\
  U_TU_S\ket{v_T}&=\ket{\Phip}_{\refq L_T}\otimes\ket{\beta},
\end{align}
with $\ket{\alpha}$, $\ket{\beta}$ unit vectors on the remaining systems.
Writing $\ket{\alpha}=\sum_i\ket{i}_{L_T}\ket{\alpha_i}$,
$\ket{\beta}=\sum_i\ket{i}_{L_S}\ket{\beta_i}$ and
$\ket{\Phip}=2^{-1/2}\sum_a\ket{aa}$, the basis labels on $\refq$, $L_S$
and $L_T$ must agree, so
\begin{equation}
  |\braket{v_S}{v_T}|
  =\frac12\Big|\sum_{i=0}^{1}\braket{\alpha_i}{\beta_i}\Big|
  \le\frac12\sum_i\|\alpha_i\|\,\|\beta_i\|\le\frac12
  \label{eq:overlap}
\end{equation}
by the Cauchy--Schwarz inequality, using
$\sum_i\|\alpha_i\|^2=\sum_i\|\beta_i\|^2=1$. Here $\ell\ge 2k$ guarantees that disjoint groups exist; the Kneser-graph bound below also assumes $\ell\ge 2k$.

\emph{Kneser-graph bound.}  Ahmed and Smith~\cite{AhmedSmith2026}
represent the $k$-subsets $S\ksub[\ell]$ as the vertices of the Kneser graph
$\KG$, two vertices being adjacent iff $S\cap T=\emptyset$.  A
$\mu$-orthonormal representation assigns to every vertex a unit vector
$\ket{v_S}$ with $|\braket{v_S}{v_T}|\le\mu$ for adjacent $S,T$, together
with a unit vector $\ket{\psi}$, and their Corollary~16 [Eq.~(S1.10)], together with their definition of the $\mu$-approximate Lovász theta function [Eq.~(2)],
states that for any such representation and any $\mu\in[0,1]$
\begin{equation}
  \sum_{S\ksub[\ell]}|\braket{\psi}{v_S}|^2
  \le\binom{\ell-1}{k-1}\Big(1+\mu\,\frac{\ell-k}{k}\Big).
  \label{eq:kneser}
\end{equation}
By \eqref{eq:overlap}, our vectors form such a representation with
$\mu=\tfrac12$.  Using \eqref{eq:FS}, \eqref{eq:handle} and
$\binom{\ell-1}{k-1}\big/\binom{\ell}{k}=k/\ell$,   
\begin{equation}
  \Fe=\frac{\sum_{S\ksub[\ell]}|\braket{\psi}{v_S}|^2}{\binom{\ell}{k}}
  \le\frac{k}{\ell}+\frac12\Big(1-\frac{k}{\ell}\Big)
  =1-\frac{\gamma}{2}.
\end{equation}

\emph{Irrational $\gamma$.}  Adding damping at the receiver turns a code
for $\AD$ into a code for $\mathcal{N}_{\gamma'}$, $\gamma'\le\gamma$, with
the same fidelity, so the optimal fidelity is non-increasing in $\gamma$;
applying the bound to rational $\gamma'\in[\tfrac12,\gamma)$ with
$\gamma'\to\gamma$ gives it for all $\gamma\ge\tfrac12$.

\emph{Tightness.}  One use with one classical bit attains $\Fe=1-\gamma/2$ (Sec.~S5.1, Eq.~\eqref{eqS:onebit_fidelity}), which corresponds to the average fidelity $\bar F_1$ of Eq.~(3) of the Letter through $\bar F=(2\Fe+1)/3$. Hence, for $\gamma\ge\tfrac12$,
additional channel uses cannot raise the entanglement fidelity of
FCA codes above the one-use value.